\documentclass{aa}
\usepackage{graphicx}
\usepackage{txfonts}
\begin{document}
\title{The enigmatic B[e]-star Henize 2-90:
The non-spherical mass loss history from an analysis of forbidden lines\thanks{Based on observations done with the 1.52m telescope at the
European Southern Observatory (La Silla, Chile), under the agreement with the
Observat\'orio Nacional-MCT  (Brazil)} \thanks{Table\,\ref{Table_1} is only available
in the online version of the journal}}

\author{M. Kraus\inst{1},  M. Borges Fernandes\inst{1,2}, F.X. de 
          Ara\'ujo\inst{2}
          \and
          H.J.G.L.M. Lamers\inst{1}
          }

\offprints{M. Kraus}
\mail{M.Kraus@phys.uu.nl}

\institute{Astronomical Institute, Utrecht University,
             Princetonplein 5, NL 3584 CC Utrecht, The Netherlands\\
             \email{M.Kraus@phys.uu.nl}
             \email{lamers@astro.uu.nl}
         \and
             Observat\'orio Nacional-MCT, Rua General Jos\'e Cristino 77,
             20921-400 S\~ao Cristov\~ao, Rio de Janeiro, Brazil\\
             \email{borges@on.br}
             \email{araujo@on.br}
             }

\date{Received xxxxxxx xx, xxxx; accepted xxxxxx xx, xxxx}

\abstract{
We study the optical spectrum of the exciting B[e] star \object{Hen 2-90} 
based on new high-resolution observations that cover the innermost
2\arcsec of the object whose total extent is more than 3\arcsec. 
Our investigation is splitted in two parts, a 
qualitative study of the presence of the numerous emission lines and the 
classification of their line profiles which indicate a circumstellar environment of 
high complexity, and a quantitative analysis of numerous forbidden lines, e.g.
[O{\sc i}], [O{\sc ii}], [O{\sc iii}], [S{\sc ii}], [S{\sc iii}], [Ar{\sc iii}], 
[Cl{\sc ii}], [Cl{\sc iii}] and [N{\sc ii}]. 
We find a correlation between the different ionization states of the elements and
the velocities derived from the line profiles: the highly ionized atoms have the 
highest outflow velocity while the neutral lines have the lowest outflow velocity. 
The recent HST image of Hen 2-90 (Sahai et al. \cite{Sahai02}) reveals a bipolar, 
highly ionized region, a neutral disk-like structure and an intermediate region of 
moderate ionization. This HST image covers about the same innermost regions as our 
observations. When combining the velocity information with the HST image of Hen 
2-90 it seems that a non-spherical stellar wind model is a good option to
explain the ionization and spatial distribution of the circumstellar material. 
Such a wind might expand into the cavity formed during the AGB phase of the star 
which is still visible as a large nebula, seen 
e.g. on H$\alpha$ plates. We modelled the forbidden lines under the assumption of 
a non-spherically symmetric wind that can be split into a polar, a disk forming 
and an intermediate wind, based on the HST image. We find that in order to fit the 
observed line luminosities, the mass flux, surface temperature, and terminal
wind velocities need to be latitude dependent, which might be explained in terms
of a rapidly rotating central star. A rotation speed of 75--80\% of the critical velocity
has been derived from the terminal velocities, extracted from the observed line wings
considering the inclination of the system as suggested from the HST image. 
The total mass loss rate of the star was determined to 
be of order $3\times 10^{-5}$\,M$_{\odot}$yr$^{-1}$. Such a wind scenario and the 
fact that compared to solar abundances C, O, and N seem to be underabundant while 
S, Ar and Cl have solar abundances, might be explained in terms of a
rapidly rotating post-AGB star. 

   }
\authorrunning{M. Kraus et al.}
\titlerunning{The enigmatic B[e]-star Hen 2-90}
\maketitle
%

\section{Introduction}

Hen 2-90 (PN Sa 2-90, PN G305.1+01.4, IRAS 13064-6103) is a very interesting object
whose evolutionary stage is unclear. Its distance was estimated to be about 
$d \, = 1.5$ kpc by Costa et al. (\cite{Costa}) while Sahai \& Nyman 
(\cite{Sahai00}) adopted $d \, = 2.5$ kpc. Its effective temperature was estimated
to be $51000$ K (Kaler \& Jacoby, \cite{Kaler}) and  its luminosity was derived 
to be $\log(L/L_{\sun}) \sim 3.0$ (Costa et al., \cite{Costa}).

The star was firstly classified as a planetary nebula by Webster (\cite{Webster}) 
and Henize (\cite{Henize}). Later, Costa et al. (\cite{Costa}) and Maciel 
(\cite{Maciel}) classified it as a planetary nebula with low metal abundance and 
with a central star of low mass and low luminosity. Lamers et al. (\cite{Lamers98}) 
included it in the list of objects presenting the B[e] phenomenon, as a compact 
planetary nebula B[e]\footnote{Note, that the classification as a B star
is based on the emission spectrum but it does not reflect the effective temperature
of the star.}. 

Sahai \& Nyman (\cite{Sahai00}), based on HST images, and Guerrero et al. 
(\cite{Guerrero}), using ground based images, have described the presence of a 
nebula bisected by a disk and with a bipolar jet and knots, spaced uniformly. 
Guerrero et al. (\cite{Guerrero}) noted that the dynamical stability of the 
jets and knots makes Hen 2-90 a unique object. These characteristics could point to a 
compact planetary nebula, since many planetary nebulae present a bipolar nebula. 
This fact implies that the asymmetries in the wind can already take place in the 
late stages of the AGB phase, where the jets can probably shape the spherical AGB 
wind in a bipolar scenario (Sahai \& Trauger, \cite{Sahai98}; Imai et al., 
\cite{Imai}; Vinkovi\'c et al., \cite{Vinkovic}). However, the presence of a disk 
and jets could also be explained assuming that the system is a binary (Sahai 
\& Nyman, \cite{Sahai00}). Guerrero et al. (\cite{Guerrero}) also 
noted, that based on near-IR colors, Hen 2-90 might be a symbiotic system, i.e.
a binary composed of a cool giant and a hot component with an accretion disk. 
Due to these two completely different interpretations (compact planetary nebula or 
symbiotic object) the real nature of Hen 2-90 stays unclear.

We study this object by means of high-resolution spectroscopy and low resolution 
spectrophotometry data obtained with the FEROS and B\&C spectrographs, 
respectively, at the 1.52m telescope in ESO (La Silla, Chile). The goals of this 
study are: 

\begin{itemize}

\item to describe the spectral features in the optical spectrum, obtained with 
a resolution higher than any other published observation to date; 

\item to study the circumstellar material of this star (temperature and density 
distribution, velocities, ionization structure) using the spectrophotometrically
calibrated line profiles. 

\end{itemize}

For the first goal, we will describe the various line profiles and associated 
Doppler velocities of the different types of lines. For the second one, we will 
make a complete analysis of all the forbidden emission lines (except of lines of
Fe{\sc ii}).

The structure of the paper is as follows.
In Sect. 2 we give the information about our observations. 
In Sect. 3 we present a spectral atlas 
obtained from our optical data, showing the different line profiles 
identified in the spectra. 
In Sect. 4 we discuss the nature of the circumstellar material whether it is 
a shell, nebula or a stellar wind.
In Sect. 5 we derive the total mass loss rate of the star from modeling the 
forbidden emission lines in the scenario of a non-spherically symmetric wind.
In Sect. 6 we discuss the validity of our assumptions and discuss the 
non-spherical wind scenario in terms of a rapidly rotating star. 
In addition, we discuss the nature of Hen 2-90, as being either a compact
planetary nebula or a symbiotic object, based on abundance 
criteria found from the analysis and on spectral characteristics in our 
new high-resolution observations. In Sect. 7 we summarize our conclusions.


\section{Observations \& Reductions}\label{observations}

High and low resolution observations were obtained with the Fiber-fed 
Extended Range Optical Spectrograph (FEROS) and with the Boller \& Chivens 
(B\&C) spectrograph, respectively, at the ESO 1.52m telescope in La Silla 
(Chile).  FEROS is a bench-mounted Echelle spectrograph with the fibers 
located at the Cassegrain focus with a spectral resolution of R = 48\,000 
corresponding to 2.2 pixels of 15\,$\mu$m and with a wavelength coverage 
from 3600\,\AA \ to 9200\,\AA. The aperture of FEROS through which the fibers
are illuminated is 2$\arcsec$.

High resolution observations of Hen 2-90 (higher than the spectra previously 
described in the literature) were taken on 2000 
June 10, with an exposure time of 4800 seconds. These data were used 
to get a spectral atlas, a detailed description 
of the line profiles and also to deblend some features present in the 
low resolution spectra. The S/N ratio in the 5500\,\AA \ region is aproximately 
20.  FEROS has a complete automatic on-line 
reduction, that was adopted by us. Equivalent widths have been
measured using an IRAF task that 
computes the line area above the adopted continuum. 

Low resolution spectra of Hen 2-90 were taken on 2000 June 9, with an 
exposure time of 900 seconds. The slit width was 2$\arcsec$. The instrumental 
setup employed made use of grating \#23 with 600 l mm$^{-1}$, providing a 
resolution of $\sim 4.6$\,\AA \ in the range 3800-8700\,\AA. The 
efficiency of the CCD is a function of wavelength. It increases from about
70\% at 3500\,\AA~to its maximum value of 90\% at about 7000\,\AA~and then 
decreases again, reaching 75\% at 8000\,\AA. This behaviour results in larger 
flux uncertainties of about 20\% especially in the very low ($< 5000$\AA) and 
very high ($> 7500$\AA) wavelength regions of the spectrum, while in the range 
in between the flux uncertainties are more of order 10\%.
In the 5500\,\AA \ region, the S/N ratio in the 
continuum is aproximately 40 for the B\&C spectrum (hereafter Cassegrain 
spectrum). Since there is no completely line free region in the spectrum, the 
S/N derived is an upper limit. The Cassegrain spectra were reduced using standard 
IRAF tasks, such as bias subtraction, flat-field normalization, and wavelength 
calibration. We have done flux calibrations, and extinction corrections with 
E(B-V) = 1.3 (Costa et al., \cite{Costa}). Spectrophotometric standards from 
Hamuy et al. (\cite{Hamuy}) were used for absolute flux calibration. In the 
linearized spectra the line intensities have been measured by the conventional 
method of adjusting a gaussian function to the profile. Another source of uncertainty 
in the line intensities is the position of the underlying continuum and we 
estimate the errors to be about 20\% for the weakest lines (line fluxes $\approx$ 
10 on a scale with H$\beta$ = 100) and about 10\% for the strongest lines.


\section{The Spectral Atlas}\label{atlas}

\begin{figure}
\resizebox{\hsize}{!}{\includegraphics{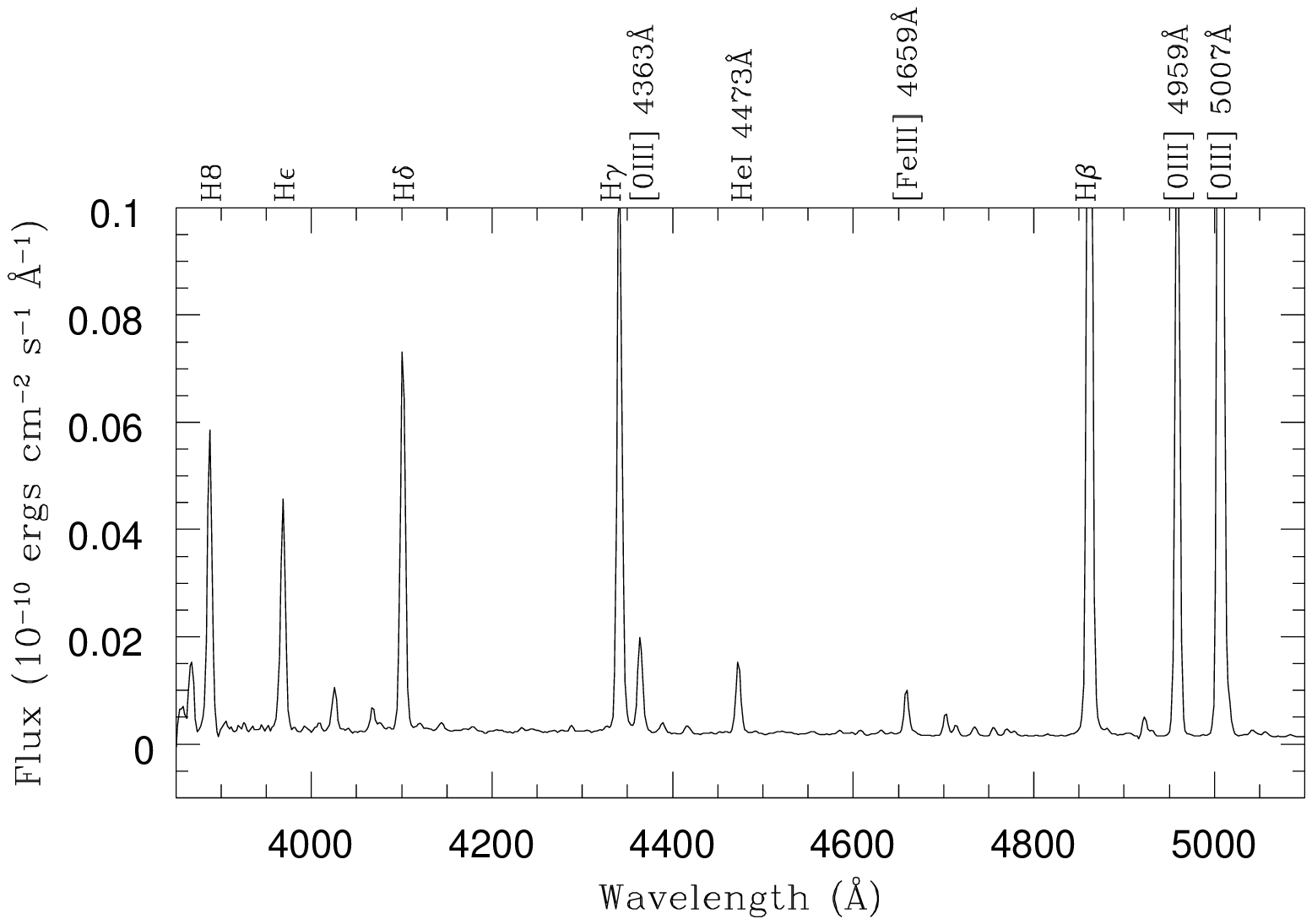}}
\vskip0.2truecm
\resizebox{\hsize}{!}{\includegraphics{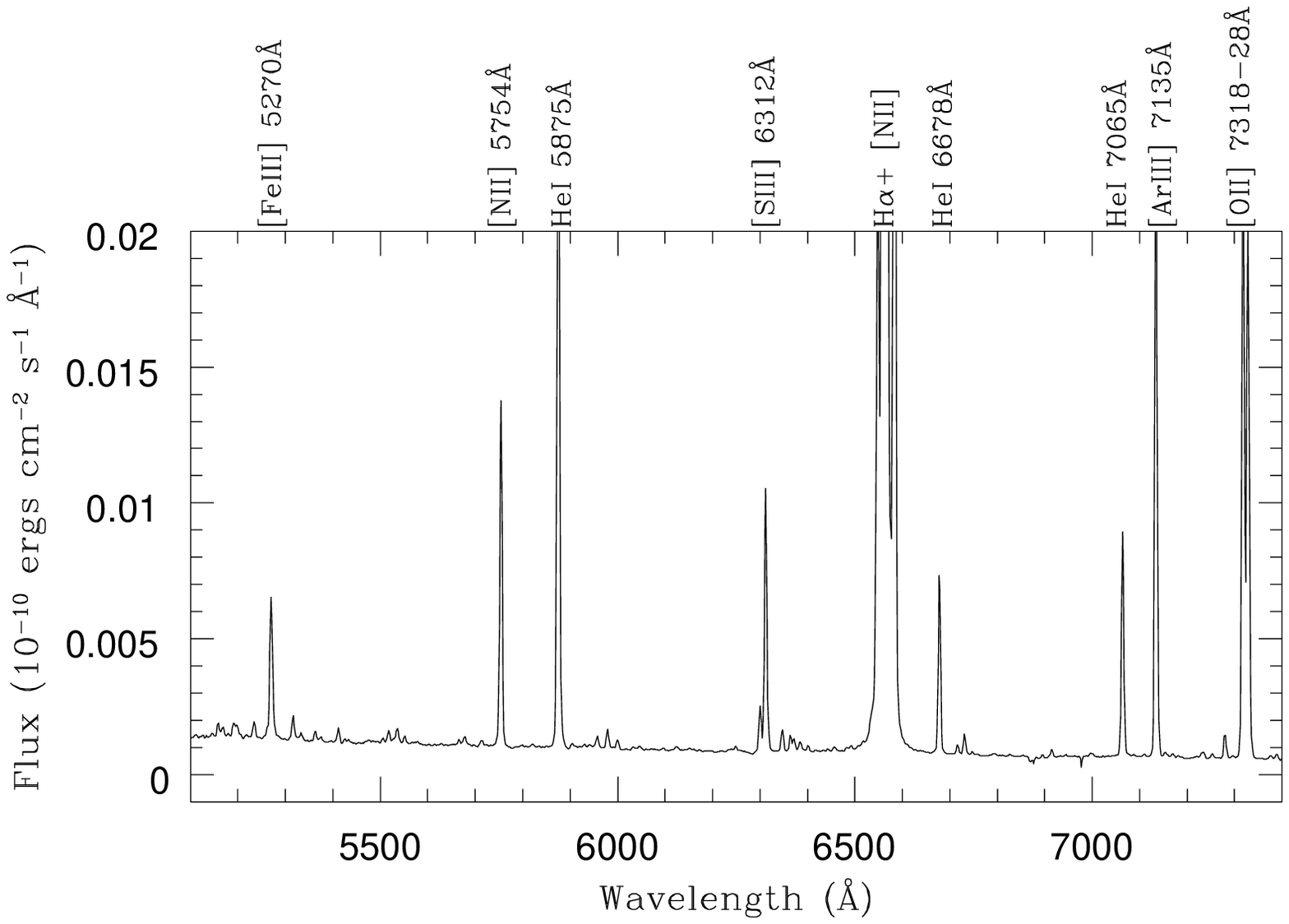}}
\caption{Extinction corrected optical spectrum of Hen 2-90, observed with the B\&C spectrograph.}
\label{Fig.1}
\end{figure}

In order to identify the lines and make a spectral atlas of the optical region,
we have used the line lists provided by
Moore (\cite{Moore}), Thackeray (\cite{Thackeray}), and Landaberry, Pereira \& Ara\'ujo (\cite{Landaberry}). We have
also looked up two sites on the web: NIST Atomic Spectra Database Lines Form
(URL physics.nist.gov/cgi-bin/AtData/lines\_form) and The Atomic
Line List v2.04 (URL www.pa.uky.edu/~peter/atomic/).

Fig.\,\ref{Fig.1} shows the extinction corrected low resolution Cassegrain spectrum and we can see that it
is dominated by emission lines superimposed on a flat continuum. No absorption line
is present. Table\,\ref{Table_1} which is only available in the online version
of the journal shows many emission lines that we have identified. There, the observed 
wavelength (Column 1), the observed intensity
(I$_{\rm obs}$($\lambda$), Column 2), the extinction corrected intensity (I$_{\rm 
corr}$($\lambda$), Column 3) and the proposed identification (Column 4) for each
line are given. The intensities are relative to H$\beta$ = 100 with an observed 
H$\beta$ flux of $2.18 \times 10^{-12}$ ergs cm$^{-2}$ s$^{-1}$ and an extinction 
corrected H$\beta$ flux of $1.35 \times 10^{-10}$ ergs cm$^{-2}$ s$^{-1}$.
The line identification given in Column\,4 of Table\,\ref{Table_1} encloses
the ionization state of the element with the proposed transition and multiplet
as well as the rest wavelength of the transition. It is possible
that more than one ion can be allocated to a single feature. In these
cases, we give some possible alternative identifications. For some lines no
identification could be found. These lines remain unidentified,
labelled as ''Uid`` in the Table.

Many permitted and forbidden lines were identified, most of them coming from
singly or doubly ionized elements.  Iron is by far the element with the
richest spectrum and [Fe\,{\sc iii}] lines are the strongest ones of this element, as 
cited by Guerrero et al. (\cite{Guerrero}). On the other hand our data show an even 
larger number of permitted and forbidden lines of Fe\,{\sc ii} (although less intense 
than the [Fe\,{\sc iii}] lines), not reported by Guerrero et al. (\cite{Guerrero}). 

The presence of very intense H\,{\sc i} Balmer lines, He\,{\sc i}, 
[O\,{\sc ii}], [O\,{\sc iii}], [N\,{\sc ii}] and [S\,{\sc iii}]
lines is remarkable. H$\alpha$ is the most intense line, while [O\,{\sc iii}] 
$\lambda$5007 is the second one. This spectral characteristic is very 
curious, because it is different from that usually seen in a typical 
planetary nebulae, where the [O\,{\sc iii}] $\lambda$ 5007 is the most intense 
line. It also differs from the spectrum of a low-excitation planetary nebulae, 
where the H$\alpha$, H$\beta$ and [N\,{\sc ii}] $\lambda$$\lambda$6548,6583 are 
more intense than [O\,{\sc iii}] $\lambda$5007 (Kwok, \cite{Kwok}). Guerrero et al. 
(\cite{Guerrero}) could neither confirm nor deny the presence of TiO absorption 
bands, the main signature of a symbiotic system, coming from the cool component. 
However, our spectrum shows clearly that the TiO bands are not present. In addition, 
we also could not identify the He\,{\sc ii} lines that would come from a hot component.

The identification of many Fe\,{\sc ii} permitted and forbidden lines, H\,{\sc i} 
Balmer emission lines and also [O\,{\sc i}] lines confirms the presence of the 
B[e] phenomenon, described by Lamers et al. (\cite{Lamers98}).

\begin{figure}
\resizebox{\hsize}{!}{\includegraphics{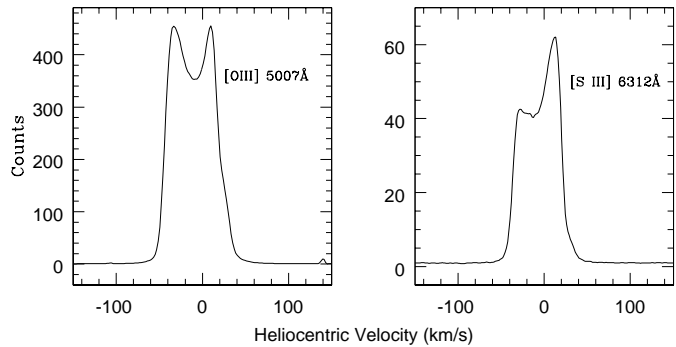}}
\caption{The shown [O\,{\sc iii}] and [S\,{\sc iii}] lines taken from the Hen
2-90 high-resolution (FEROS) spectrum are only two examples for the presence of
double-peaked emission line profiles. The velocities obtained from the wings of 
these lines are given in Table\,\ref{Table_2}. Due to the strong lines, the high 
velocity wings are not resolved in this figure.}
\label{Fig.2}
\end{figure}

\begin{figure}
\resizebox{\hsize}{!}{\includegraphics{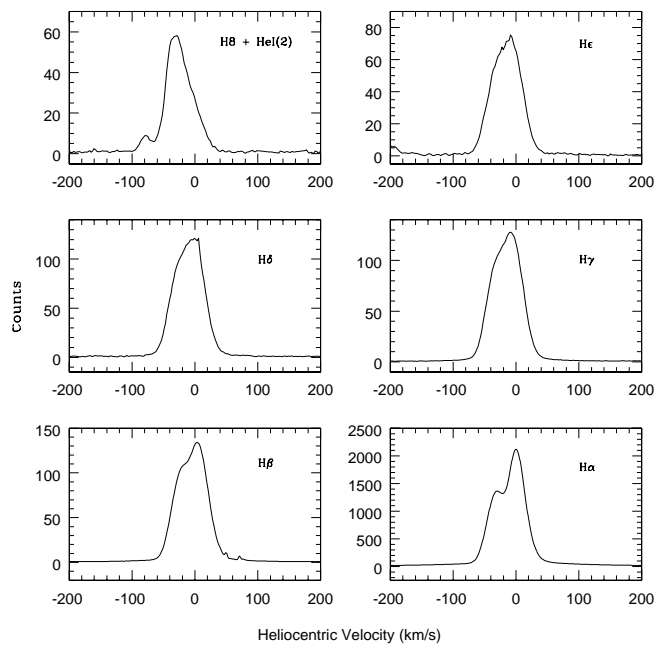}}
\caption{The H\,{\sc i} Balmer lines taken from the Hen 2-90 high-resolution 
(FEROS) spectrum. These lines show an evolution from single-peaked profiles
over profiles with a shoulder to double-peaked profiles with decreasing quantum
number.}
\label{Fig.3}
\end{figure}


\begin{figure}
\centering
\resizebox{\hsize}{!}{\includegraphics{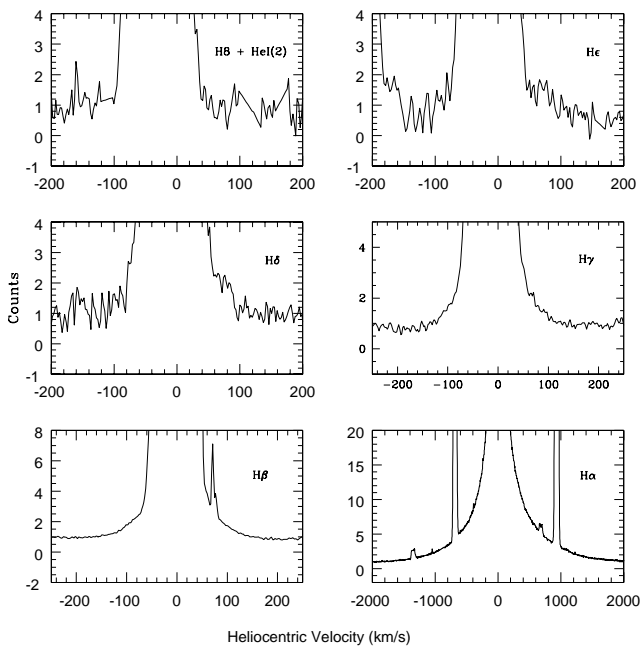}}
\caption{Wings of the Balmer lines taken from the Hen 2-90 high-resolution (FEROS)
spectrum. They show an increase in velocity from H8 to H$\alpha$ (see 
Table\,\ref{Table_2}). The wings of H$\alpha$ extend to velocities around 1800 km 
s$^{-1}$; superimposed are the [N\,{\sc ii}] lines ($\lambda$$\lambda$6548,6583) 
as well as the only carbon line identified (C{\sc ii}, $\lambda$6578).}
\label{wings}
\end{figure}

\begin{figure}
\resizebox{\hsize}{!}{\includegraphics{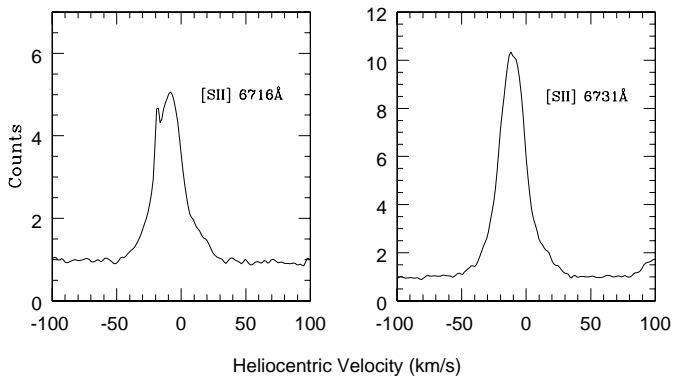}}
\caption{[S\,{\sc ii}] line profiles present in the high-resolution spectrum
(FEROS) of Hen 2-90. The derived wing velocities are given in Table\,\ref{Table_2}.}
\label{Fig.5}
\end{figure}

\begin{figure}
\resizebox{\hsize}{!}{\includegraphics{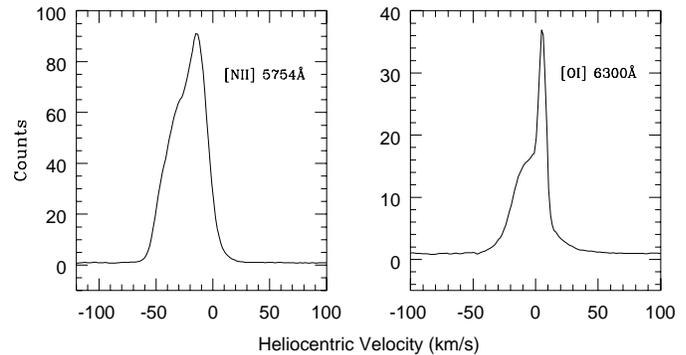}}
\caption{The [N\,{\sc ii}] $\lambda$5754 and the [O\,{\sc i}] $\lambda$6300 
lines present in the Hen 2-90 high-resolution (FEROS) spectrum. Notice
the presence of a "shoulder" or almost a second peak. The velocities 
obtained from the wings of these lines are given in Table\,\ref{Table_2}.}
\label{Fig.6}
\end{figure}

Concerning the line profiles, Guerrero et al. (\cite{Guerrero}) grouped the emission 
lines of Hen 2-90 into three groups: lines with (1) double peaks, (2) broad single 
peaks (FWHM $\geq$ 40 km s$^{-1}$), and (3) narrow single peaks (FWHM $\leq$ 
30 km s$^{-1}$). Using our FEROS spectrum, we confirm the presence 
of those groups and define a new one, (4) the lines that show clearly a "shoulder", 
an intermediate case between the double peaks and the single peak. 
We shortly list the lines falling into each profile group and show an example:

{\bf (1) Double peaked profiles} are shown by the forbidden lines of O\,{\sc ii}, 
O\,{\sc iii} (left panel in Fig.\,\ref{Fig.2}), Ar\,{\sc iii}, Fe\,{\sc iii}, S\,{\sc 
iii} (right panel in Fig.\,\ref{Fig.2}), and Cl\,{\sc iii} with a peak separation of 
$\sim$ 40 km s$^{-1}$. 
Also H$\alpha$ shows a double peak structure (Fig.\,\ref{Fig.3}). 
It is interesting to note that the asymmetric double-peaked lines have 
the red peak more intense than the blue one, indicating 
that the receding material is brighter than the approaching one.
This finding is in agreement with Guerrero et al. (\cite{Guerrero}). However, contrary
to these authors, our [O\,{\sc iii}] lines (except of the 4363\,\AA~line) show 
double peaks with equal strength. The difference between the line profiles (and also 
the line intensities) are probably due to the different slit widths used for the 
observations.

The double-peaked forbidden lines show, in general, emission wings extending from 45 
to around 100 km s$^{-1}$ (except of H$\alpha$), indicating the presence of a low 
velocity wind. H$\alpha$ however, shows wings extending up to 1800 km s$^{-1}$ 
(Fig.\,\ref{wings}). Whether these wings indicate an outflow velocity or are produced due to 
electron scattering will be discussed in Sect.\,\ref{balmerwings}. 

The velocity extent of the H$\alpha$ wings, in our data, is higher than the
1050\,km\,s$^{-1}$ cited by Costa et al. (\cite{Costa}), and the 1500\,km\,s$^{-1}$
found by Guerrero et al. (\cite{Guerrero}). The differences in these measurements
are due to the different resolutions of the spectra and due to the different
slit widths. However, only more observations with similar instrumental setup can 
discard a spectral variation. 

Another difference between our spectra and the data presented by Guerrero et al. 
(\cite{Guerrero}) and worth mentioning, is the absence in their spectra of the 
[S\,{\sc iii}] $\lambda$6312 emission. This line is very strong in our spectrum and 
shows double-peaks (see Fig.\,\ref{Fig.2}). A possible explanation might be that since 
this emission line arises in the outer parts of the intermediate wind (at least 
according to our non-spherical wind model, see Sect.\,\ref{results} and 
Fig.\,\ref{sulfur3}), the slit width used by Guerrero et al. (\cite{Guerrero}) was 
too narrow to detect this line.

{\bf (2) Broad single peaked profiles} are found for H$\delta$, 
H$\epsilon$, H8 (Fig.\,\ref{Fig.3}), Paschen lines, and some He\,{\sc i} lines. 
The FWHM of these lines is typically $\geq$ 40 km s$^{-1}$.

{\bf (3) Narrow single peaked profiles} are shown especially by the forbidden lines
of Fe\,{\sc ii} and S\,{\sc ii} (Fig.\,\ref{Fig.5}). These lines typically 
have FWHM $\leq$ 30 km s$^{-1}$.

{\bf (4) Profiles with a "shoulder"} or almost a second peak which are in the 
new group identified by us. This group is represented by some He\,{\sc i}, [N\,{\sc 
ii}], and [O\,{\sc i}] lines (Fig.\,\ref{Fig.6}), as well as by H$\beta$ and H$\gamma$ 
(Fig.\,\ref{Fig.3}). These lines are asymmetrical, and in some cases their blue 
and red wings show different velocities (see Table\,\ref{Table_2}).

It is interesting to note that the H\,{\sc i} Balmer lines show an evolution 
from a single peak in the high order lines to double peak in H$\alpha$, having
an intermediate case with shoulder in H$\beta$ and H$\gamma$ (Fig.\,\ref{Fig.3}). 

Table 2 shows the velocities derived from the maximum extend of the wings of 
several forbidden emission lines and some Balmer lines in our FEROS spectrum. There, 
the ion identification (Column 1), laboratory wavelength ($\lambda$ (\AA), Column 2), blue wing 
velocity (v$_{\rm blue}$ (km s$^{-1}$), Column 3) and red wing velocity (v$_{\rm red}$ 
(km s$^{-1}$), Column 4) for each line are given. Included are only lines whose 
velocities could be determined with high accuracy and which are not blended.

\begin{table}
\caption{Derived maximum velocities where the blue and red wings reach the continuum.
Listed are some forbidden emission lines and some Balmer lines in the FEROS spectrum whose
maximum wing velocities could well be determined and which are not blended.} 
\begin{tabular}{ccrr}
\hline
\hline
Ion & $\lambda$ &   v$_{\rm blue}$ & v$_{\rm red}$ \\
    &  [\AA] &  [km s$^{-1}$]  &  [km s$^{-1}$] \\
\hline
O{\sc iii}  & 4363 & -75 & +45 \\
O{\sc iii}  & 4959 & -80 & +65  \\
O{\sc iii}  & 5007 & -70 & +60 \\
O{\sc i}    & 5577 & -25 & +25 \\
O{\sc i}    & 6300 & -50 & +40 \\
O{\sc i}    & 6364 & -50 & +30 \\
S{\sc iii}  & 6312 & -60 & +50 \\
S{\sc ii}   & 4068 & -25 & +25 \\
S{\sc ii}   & 4076 & -50 & +35 \\
S{\sc ii}   & 6716 & -50 & +30 \\
S{\sc ii}   & 6731 & -50 & +35 \\
N{\sc ii}   & 5755 & -70 & +20 \\
N{\sc ii}   & 6548 & -70 & +45 \\
N{\sc ii}   & 6584 & -80 & +50 \\
Fe{\sc ii}  & 7155 & -40 & +30 \\
Cl{\sc iii} & 5517 & -30 & +100 \\
Cl{\sc iii} & 5538 & -50 & +40 \\
H$\alpha$   & 6563 & -1800 & +1800 \\
H$\beta$    & 4861 & -160 & +160 \\
H$\gamma$   & 4340 & -140 & +140 \\
H$\delta$   & 4101 & -100 & +100 \\
H$\epsilon$ & 3970 & -100 & +100 \\
H8          & 3889 & -100 & +40 \\
\hline
\end{tabular}
\label{Table_2}
\end{table}

In summary, we can say that compared with the Guerrero et al. (\cite{Guerrero})
data, our spectra revealed many more emission lines due to the higher signal-to-noise
ratio (see Table\,\ref{Table_1}).
The fact that with the same applied reddening correction value some of the line
profiles and intensities seem to have changed over a period of 6 months needs some
further clarification. Our line profiles shown have been taken
with FEROS making use of the fiber technique and with an aperture of  
2$\arcsec$. A slit width of 2$\arcsec$ was also used for the Cassegrain spectrum
from which we derived the line intensities. 
Guerrero et al. (\cite{Guerrero}) used a slit width of 1$\arcsec$, i.e.
only half of our value. They observed therefore a region very close to the star,
while our spectra cover a much larger region of the circumstellar material.
In addition, the dispersion and signal-to-noise ratio of the spectra are different,
which makes it difficult to compare individual lines. The question of line variability
can therefore only be answered by subsequent observations with identical setups,
and differences between the Guerrero et al. (\cite{Guerrero}) spectrum and our data
should not be interpreted as source variability. What we can conclude, however,
is that the circumstellar medium around Hen 2-90 is not homogeneous. On the contrary,
the huge zoo of observed emission lines gives us a clear hint that the circumstellar
medium must have a rather complex structure in both, density and temperature,
as indicated by the variety of line profiles, velocities and ionization states.


\section{The nature of the circumstellar material: nebula or wind?}

In this section we discuss the nature of the circumstellar material close
to the central star, whether it 
can be described by the nebula approximation often used for the analysis of planetary
nebula spectra or whether we have to apply a wind scenario. This discussion is based
on the available observations: our optical spectra in combination with the HST image
published by Sahai et al. (\cite{Sahai02}, see top panel of Fig.\,\ref{HST}).

{\bf Emission lines:} We observe from the innermost regions in Hen 2-90 permitted 
emission lines as well as forbidden emission lines. The huge amount of forbidden lines
might speak in favour of a nebula nature of the emitting material. However, we cannot 
explain the evenly huge amount of permitted lines with the nebula approximation 
because both sorts of lines need completely different density conditions: permitted 
lines are produced in regions of high density, forbidden lines are produced in regions 
of low density. The fact that both types of lines are very prominent in our spectra 
indicates that we need at least two different regions, one with high and one with low 
density. This might be consistent with the picture of a high density (ring?) nebula 
in the equatorial region and a low density nebula in polar region.

{\bf Ionization structure:} Inspection of the HST image (Fig.\,\ref{HST}) shows
clearly high-ionized material in the polar direction (= jet direction in the 
image), i.e. in the presumably low-density 
region, dominated by emission of [O{\sc iii}], and low-ionized material at more 
intermediate latitudes (counted from the pole), dominated by [N{\sc ii}] emission. 
In our spectra, we found many forbidden emission lines as listed in 
Table\,\ref{Table_1}. These lines are from ions in different ionization states 
and must therefore arise in regions of different ionization and therefore different
temperature, consistent with the HST image. In addition, we found
very strong emission of [O{\sc i}] which means that this emission region must be 
neutral, also in hydrogen. The emission of the [O{\sc i}] lines therefore must 
arise in (i) a (hydrogen) neutral region, speaking for high density material at least
close to the star which guarantees shielding from the ionizing stellar flux, and 
(ii) in a medium of low density for the forbidden lines to be strong. The best location
therefore is the equatorial neutral disk-like structure visible in the HST image.

{\bf Velocity structure:} Our high-resolution spectra allow us to derive also the 
velocities from individual line profiles (see Table 2). We find that forbidden 
lines of different ionization stage which can be connected with the different 
ionization regions seen in the HST image have different velocities. The high-ionized 
lines thereby have highest velocities, the [O{\sc i}] lines have lowest velocity.

The combination of all these results, i.e. the co-existence of high- and low-density 
regions, a latitudal velocity structure having highest values in polar and
lowest values in equatorial directions, a latitude dependent ionization
structure of the circumstellar material which also infers a latitudal temperature 
structure, and the variety of line profiles 
described in the previous section, seems to be more consistent with a non-spherical
stellar wind scenario rather than with a simple constant density nebula scenario
or even a piece-wise constant density nebula approximation.
This conclusion has been drawn solely from the qualitative analysis of our 
observations in combination with the information that follow from the HST image.
As we will show in Sect.\,\ref{results} when modeling the line luminosities of the 
forbidden emission lines, it turns out that the saturation of the line luminosities 
happens for different lines at different distances from the star. This means, that
the forbidden lines are very sensitive to the density structure,
supporting the idea of a radial density distribution rather than a constant density.


\section{Modeling of the non-spherical wind}\label{modeling}

In the previous sections we have shown that the optical spectrum of Hen 2-90
contains many forbidden and permitted emission lines. In this section we
will analyse some of the forbidden emission lines to derive the physical parameters
of the emitting gas such as electron temperature and density distribution, mass loss 
rate of the star, and the ionization structure of the circumstellar nebula.
The stellar parameters of Hen 2-90 given by Costa et al. (\cite{Costa}) have been
derived using the spherically symmetric nebula approximation and are therefore
relatively uncertain. Due to the lack of better parameter estimates we assume for our
model calculations the star to be at a mean distance of 2\,kpc, having an effective 
temperature of about 50\,000\,K and a radius of 0.38\,R$_{\odot}$ (Kaler \& Jacoby,
\cite{Kaler}; Costa et al., \cite{Costa}).

\subsection{The wind geometry}\label{geometry}

\begin{figure}
\resizebox{\hsize}{!}{\includegraphics{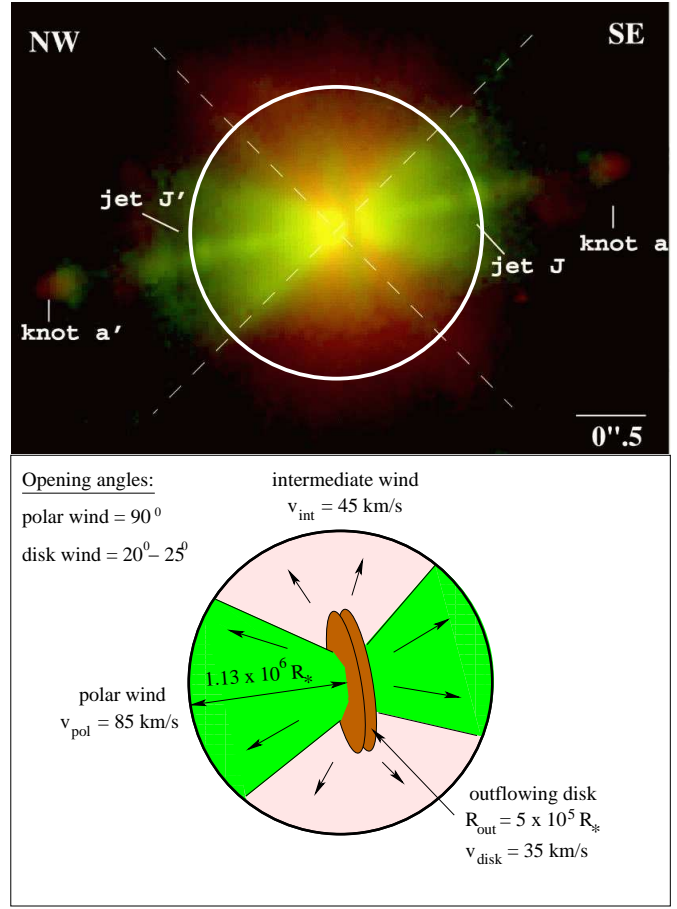}}
\caption{Top panel: HST image of Hen 2-90 replotted with kind permission from Sahai 
et al. (\cite{Sahai02}). Overplotted is a big circle that indicates 
the width of the slit (which is 2$\arcsec$, corresponding to 0.019\,pc at a 
distance of 2\,kpc) of our Cassegrain observations from which the line 
luminosities have been derived. Bottom panel: Sketch of the
threefold wind derived from the HST image and used in our calculations. Plotted 
are the outflowing disk, the polar wind, and the 
intermediate wind. Indicated are also the opening angles of the polar and disk wind 
used for the model calculations, and the velocities within the different
wind regions as derived from Table\,\ref{Table_2} and discussed in the text. 
The whole structure is still embedded in an H{\sc ii} region that extends up 
to 3$\arcsec$.2 corresponding to 0.031\,pc, which is 1.6 times larger than the 
non-spherical wind structure. The bipolar jet and the knots are not taken into 
account in the non-spherical wind model.}
\label{HST}
\end{figure}

Based on the HST image of Sahai et al. (\cite{Sahai02}, see Fig.\,\ref{HST}) we can 
distinguish three major wind regions: (i) a biconical high-ionization 
([O{\sc iii}] dominated) polar wind, (ii) a low-ionization ([N{\sc ii}] dominated) 
wind at intermediate latitudes which we will call further on the intermediate wind,
and (iii) an equatorial outflowing disk. The big white circle indicates 
the slit width of 2$\arcsec$ of our Cassegrain observations (the flux observed 
comes from this region). This corresponds to an outer edge
of about 0.01\,pc at a distance of 2\,kpc. The HST resolved structure is embedded 
into a much larger H{\sc ii} region with extensions of at least 3$\arcsec$.2 x 
3$\arcsec$.1 (Tylenda et al., \cite{Tylenda}). Our observations therefore contain 
neither much contribution of this extended nebula nor from the jets and knots, but 
are concentrated on the innermost non-spherical wind structure. 

The HST image also gives a hint to the location of the different ions. In 
combination with the velocity information retrieved from the line profiles 
(Table\,\ref{Table_2}) we can determine the terminal wind velocities of the different 
regions. For the polar wind we find a mean velocity of $\sim 60$\,km\,s$^{-1}$ 
from the [O{\sc iii}] and [Cl{\sc iii}] lines. Taking into account the opening angle 
of the cone which is about 90$\degr$ and the fact that the system is seen (almost 
perfectly) edge-on, this results in a terminal velocity of $\sim 
85$\,km\,s$^{-1}$. For the intermediate wind where the [S{\sc iii}], [S{\sc ii}] 
and [N{\sc ii}] lines come from, we derive analogously a mean wind velocity of 
$\sim 45$\,km\,s$^{-1}$, and for the disk wind, represented by the [O{\sc i}] lines, 
we find a mean velocity of about 35\,km\,s$^{-1}$.
The wind velocity shows a latitude dependence being highest in polar directions, and
lowest in equatorial direction, the difference is a factor of $\sim 2.5$.

With the adopted distance of 2\,kpc we can derive linear scales for the different
wind regions from the HST image. The polar wind is visible up to a distance of about
$1.5\times 10^{6}$\,R$_{*}$ which is slightly beyond our observing
slit which only extends to about $1.13\times 10^{6}$\,R$_{*} \simeq 0.01$\,pc. 

The outer edge of the disk is about $5\times 10^{5}$\,R$_{*}$
and we estimate, based on the dark disk-like structure seen on the 
HST image, a total opening angle of $20-25\degr$.

The remaining volume is filled by the intermediate wind which also shows at its 
innermost parts indication of a high degree of ionization ([O{\sc iii}] dominated).
We estimate that this region extends roughly to about $r_{\rm out}$(O{\sc iii})$ 
\simeq 10^{5}$\,R$_{*}$.

\subsection{The forbidden emission line luminosities}\label{lum}

We concentrate our detailed analysis on the forbidden lines.
These lines are excellent indicators of the circumstellar material because of
two reasons:\\
(i) They are excited collisionally. Therefore they are very sensitive to the
density and temperature.\\
(ii) The circumstellar nebula is optically thin for these lines, simplifying the
analysis.\\

We model the line luminosities of the forbidden emission lines of O{\sc i}, 
O{\sc ii}, O{\sc iii}, N{\sc ii}, Cl{\sc ii}, Cl{\sc iii}, Ar{\sc iii}, S{\sc ii}, 
and S{\sc iii} present in our spectra (see Table\,\ref{Table_1}). We also find 
lines from Cr{\sc ii} and Cr{\sc iv} which we could not model due to the lack of 
collision parameters, and lots of Fe{\sc ii} lines which we also neglect because 
Fe{\sc ii} cannot be treated in such an easy way as the other ions.

The line luminosity of a forbidden line is given by
\begin{equation}
L_\nu=\int j_\nu (1-W(r)) dV
\end{equation}
where $j_\nu = h\nu N_{\rm i} A_{\rm ij}$ is the line emissivity and 
$W(r) = \frac{1}{2}(1-\sqrt{1-(R_*/r)^{2}})$ is the geometrical dilution factor 
that accounts for the fraction of photons being intercepted by the star.
For the line emissivity we need to calculate the level population. This is done for
all the elements cited above by solving the statistical equilibrium equations in a
5-level atom. The collision parameters are taken from Mendoza (\cite{Mendoza}) and
the atomic parameters from Wiese et al. (\cite{Wiese1}, \cite{Wiese2}).  

The hydrogen density in a wind follows from the mass continuity equation
\begin{equation}
N_{\rm H}(r) = \frac{\dot{M}}{4\pi r^{2} \mu m_{\rm H} v(r)}
\end{equation}
$\dot{M}$ is the mass loss rate of the star and $v(r)$ the wind velocity which is given
by a $\beta$-law of the form
\begin{equation}
v(r)=v_\infty \left( 1-\frac{r_0}{r}\right)^\beta
\end{equation}
with
\begin{equation}
r_0=R_* \left\{ 1-\left(\frac{v_0}{v_\infty}\right)^{\frac{1}{\beta}} \right\}
\end{equation}
which sets the velocity at $R_*$ equal to the sound velocity ($v_0$).  $v_\infty$ is 
the terminal velocity in each wind region which has been derived above from the 
different forbidden lines, and we set $\beta = 1$ which is a good approximation for 
winds of hot stars (Lamers \& Cassinelli, \cite{Lamers99}).

The electron density is given in terms of the hydrogen density and depends on the 
degree of ionization in each part of the wind; in polar direction and in the inner
parts of the intermediate wind we set $N_{\rm e}\simeq 1.1\,N_{\rm H}$. In these
regions helium is singly ionized contributing about 10\% to the total
electron density. The contribution from other elements, like O, S, N, which are also 
(partly twice) ionized, to the total electron density is, however, negligible.
For the outer parts of the intermediate wind, where He is assumed to be neutral, we use
$N_{\rm e}\simeq 1.0\,N_{\rm H}$ (and neglect again the small contribution from 
the metals), and for the disk wind, where we assume that 
hydrogen is neutral, we have only electrons from elements with much lower
ionization potential (i.e. lower than about 10\,eV). Summing up the number 
densities of all these elements (assuming solar abundances), results in a maximum 
electron density in the disk of about $1.3\times 10^{-4}\,N_{\rm H}$. We cannot expect that all
possible elements are completely ionized. We therefore use an electron density of
$N_{\rm e}\simeq 10^{-4}\,N_{\rm H}$ which is an {\it upper limit} to the real disk 
electron density. Since the
excitation of the forbidden lines depends on the electron density, the fitting
of the disk lines therefore results in a {\it lower limit} of the mass flux,
i.e. the disk mass loss rate, because a decrease in electron density needs an increase
in total density to account for the observed line luminosities. 

\begin{figure}
\resizebox{\hsize}{!}{\includegraphics{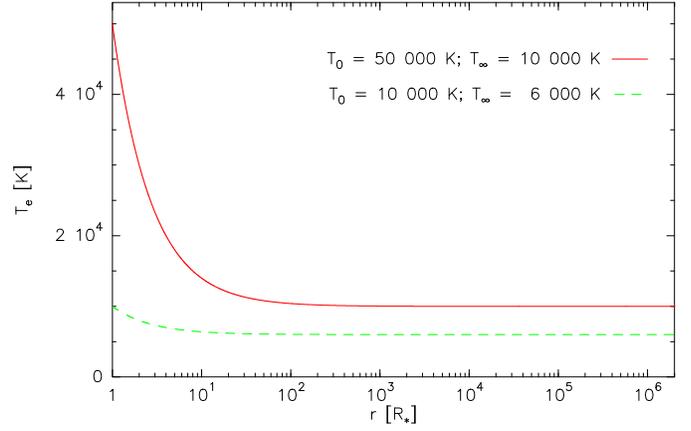}}
\caption{Temperature distribution in the wind according to Eq.\,(\ref{tempdistr}).
The starting values, $T_{0}$, and terminal values, $T_{\infty}$ appropriate
for the polar/intermediate wind and for the outflowing disk are indicated.
The temperature drops quickly (within 100\,R$_{*}$) to its terminal value which 
is found to be about 10\,000\,K in the polar and the intermediate wind, and 
about 6\,000\,K in the disk wind (see Table \ref{param}).}
\label{temp}
\end{figure}

For the temperature distribution we make use of the following equation
\begin{equation}
T(r)=T_\infty + (T_0 - T_\infty)\left(\frac{r}{R_*}\right)^{-x} \label{tempdistr}
\end{equation}
Such a temperature distribution has been found by de Koter (\cite{deKoter}) to be
a reasonable approximation for a wind of a hot star in radiative equilibrium.
We set $x = 1$, but the value of x does not really influence the results because 
the forbidden lines are formed in the lower density region, far away from the star,
where the temperature has dropped already to its terminal value, $T_{\infty}$
(see Fig.\,\ref{temp}). The 
line luminosities are, however, very sensitive to this terminal temperature. 

One additional parameter that needs to be specified is the elemental abundance. 
Costa et al. (\cite{Costa}) have found that the elements are slightly sub-solar in
Hen 2-90. However, these values have been derived under the assumption of a 
spherically symmetric homogeneous nebula which is far from being the case. In 
addition, Kraus (\cite{Kraus}) showed that the abundances when derived using a 
stellar wind rather than a homogeneous nebula can be much higher. We therefore start
our calculations assuming solar abundances (taken from Grevesse \& Sauval, 
\cite{Grevesse}) for all elements. Deviations in the calculated line luminosities 
compared with observations might then be due to individual deviations in the 
abundances.

\subsection{Mass fluxes and total mass loss rate}\label{results}

\begin{figure}
\resizebox{\hsize}{!}{\includegraphics{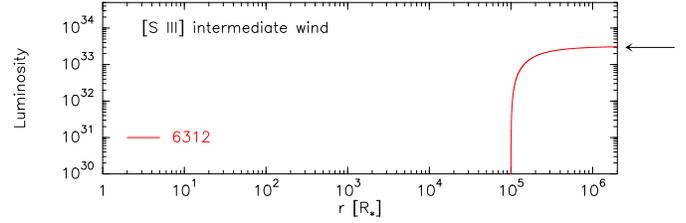}}
\caption{Model results for the [S{\sc iii}] line arising in the
outer parts of the intermediate wind.
Shown is the increase in line luminosity with distance from the star.
The arrow on the right side of the panel indicates the observed
value.}
\label{sulfur3}
\end{figure}

\begin{figure}
\resizebox{\hsize}{!}{\includegraphics{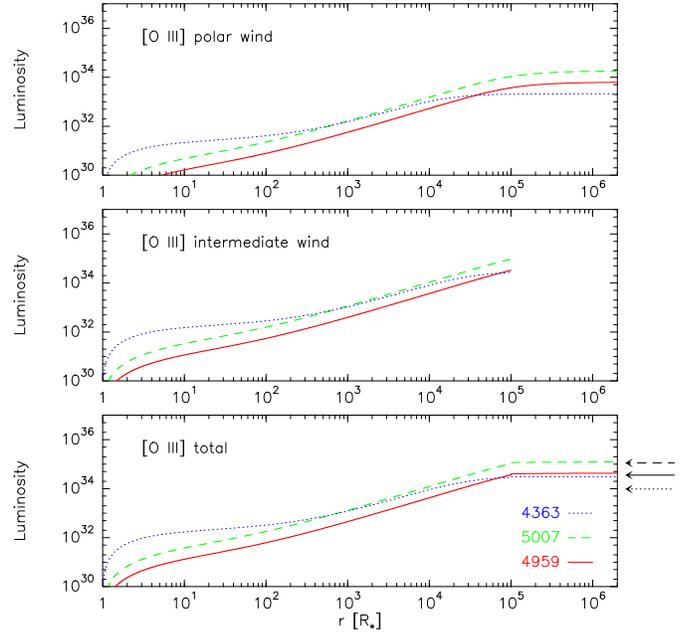}}
\caption{Model results for the different [O{\sc iii}] lines listed in the bottom
panel. Shown are the individual contributions from the polar and intermediate
wind and their line luminosity increase with distance from the star.
The bottom panel shows the sum of both contributions, and the arrows on the right side indicate the observed value for each emission line (same line style).} 
\label{oxygen3}
\end{figure}

\begin{figure}
\resizebox{\hsize}{!}{\includegraphics{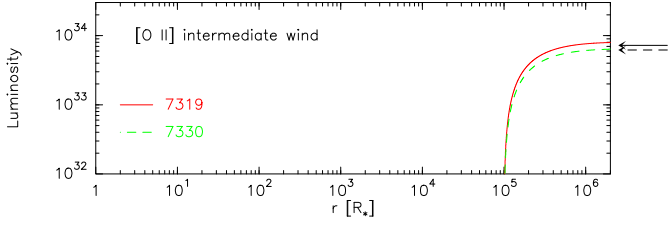}}
\caption{Model results for the [O{\sc ii}] lines.  
These lines come from the intermediate wind only, and from distances
where O{\sc iii} has already recombined (see mid-panel of Fig.\,\ref{oxygen3}).
The arrows on the right side indicate the observed
value for each emission line (same line style).}
\label{oxygen2}
\end{figure}

\begin{figure}
\resizebox{\hsize}{!}{\includegraphics{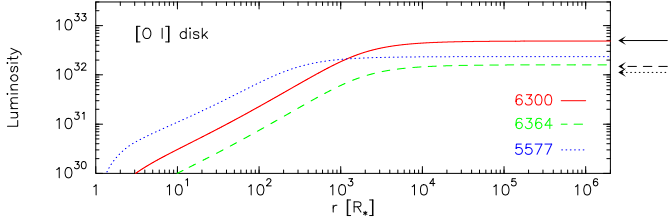}}
\caption{Model results for the [O{\sc i}] lines. 
These lines come from the disk wind only.
The arrows on the right side indicate the observed
value for each emission line (same line style).}
\label{oxygen1}
\end{figure}

The results for the ions which are used to restrict the different parameters
in our simple, non-spherical wind model are shown in Figs.\,\ref{sulfur3} - 
\ref{oxygen1}, where we plotted the line luminosities as a function of $r$, i.e. 
the volume integrated flux that is achieved in the corresponding wind (i.e. 
polar, intermediate or outflowing disk) as it would be observed from such 
a wind with outer edge $r$. For several ions, the observed line luminosities 
do not come from only one of the three defined winds, but from two different 
wind regions. Therefore we shortly describe the procedure of how we fitted 
the different lines. 

We started our modeling with the [S{\sc iii}] 6312\AA~line 
(Fig.\,\ref{sulfur3}). Since S{\sc iii}
and O{\sc ii} have about the same ionization potential, 
it can be expected that no S{\sc iii} emission will come from regions 
of O{\sc iii} emission. S{\sc iii} emission is therefore restricted to the 
intermediate wind at distances from the star, where O{\sc iii}, and hence S{\sc
iv}, have recombined already. Therefore, the inner edge of the S{\sc iii} emission
is the outer edge of the O{\sc iii} emission, i.e. $r_{\rm in}$(S{\sc iii}) $= r_{\rm
out}$(O{\sc iii}) $= 10^{5}$\,R$_{*}$ as found in Sect.\,\ref{geometry}. 

The only free parameters are the temperature distribution
and the mass loss rate at these intermediate latitudes. Since the S{\sc iii} emission
region is already rather far away from the star, we assumed the temperature to be 
constant and found a good fit for $T_{\rm e}\simeq 10\,000$\,K. The mass flux of
the intermediate wind has been found to be of order 
$F_{\rm m, int} = 1.5\times 10^{-1}$\,g\,s$^{-1}$cm$^{-2} = 1.66\times 
10^{-6}$\,M$_{\odot}$yr$^{-1}$steradian$^{-1}$. 

Since S{\sc iii} has only one prominent forbidden line in our spectral region, this 
best fit temperature had to be confirmed with the other forbidden lines from the same 
wind region. Other elements like N{\sc ii}, S{\sc ii} and Cl{\sc ii} have more than 
one forbidden line and several line ratios can be used as tracers for the electron 
temperature. From combining the results from all these temperature indicators the best 
solution was found to be of order 10\,000\,K. The valid range of temperatures is very 
small, and the resulting uncertainty in mass flux is of order 20\,\%. We want to 
stress that the dependence of the mass flux on the electron temperature cannot be
described with a power law because the collision parameters have a more complicated 
temperature dependence.

The next element to fit is oxygen and we started with the lines of O{\sc iii}.
These lines are supposed to come from 2 different regions (see Fig.\,\ref{HST} 
and Fig.\,\ref{oxygen3}): the inner parts of the intermediate wind and 
the polar wind. To get self-consistent results for all oxygen lines of 
different ionization state (Figs.\,\ref{oxygen2} and \ref{oxygen1}), we had 
to decrease the O abundance from solar to 0.3 solar. 

The contribution of the intermediate wind to the [O{\sc iii}] lines is restricted
to distances within $r_{\rm out}$(O{\sc iii}). For the temperature
distribution we used $T_{0} = T_{\rm eff}$ and $T_{\infty} = 10\,000$\,K.
The use of the effective stellar temperature as the starting temperature is quite 
reasonable because the star is supposed to be small where the surface 
temperature mirrors about the effective temperature. In fact, the choice of the 
surface temperature does not really affect the behaviour of the line luminosities 
because the majority of the luminosity comes from the lower density regions
where the temperature has reached its terminal value.

Having the contribution from the intermediate wind the 
remaining line luminosities to the [O{\sc iii}] lines 
must come from the polar wind. 
Here, the free parameters are again the mass flux and the temperature
distribution. We found best agreement by using the same temperature distribution as 
for the intermediate wind and a polar mass flux of $F_{\rm m, pol} = 7.2\times
10^{-2}$\,g\,s$^{-1}$cm$^{-2} = 7.99\times
10^{-7}$\,M$_{\odot}$yr$^{-1}$steradian$^{-1}$.  

With these fixed parameters for the intermediate wind, we found good 
agreement of observed and modelled [O{\sc ii}] lines (Fig.\,\ref{oxygen2}). 
The remaining O lines are the [O{\sc i}] lines which are 
assumed to arise in the cool and neutral equatorial disk (Fig.\,\ref{oxygen1}). 
The only constraint for the 
modeling is the temperature distribution, which should start with 
a surface temperature around 10\,000\,K to guarantee that H is neutral. The terminal 
temperature as well as the disk mass flux could then be fixed by the best fit model 
(see Fig.\,\ref{oxygen1}) to $T_{\infty} = 6\,000$\,K and 
$F_{\rm m, disk} = 5.5\times 10^{-1}$\,g\,s$^{-1}$cm$^{-2} = 6.10\times
10^{-6}$\,M$_{\odot}$yr$^{-1}$steradian$^{-1}$.

\begin{table}
\caption{Best fit model parameters for the polar (p), intermediate (i) and disk wind
(d). The mass fluxes are assumed to be constant within the indicated $\theta$-ranges
and have an error of order 20\,\%.}
\begin{tabular}{lrrrccc}
\hline
\hline
 & $T_{0}$  &  $T_{\infty}$ & $\theta_{\rm min}$ & $\theta_{\rm max}$ &
   $v_{\infty}$ & $F_{\rm m}(\theta)$ \\
 & [K] & [K] & [$\degr$] & [$\degr$] & [km\,s$^{-1}$] &
   [g\,s$^{-1}$cm$^{-2}$] \\
\hline
p & 50\,000 & 10\,000 &  0 & 45 & 85 & $7.2\times 10^{-2}$ \\
i & 50\,000 & 10\,000 & 45 & 78 & 45 & $1.5\times 10^{-1}$ \\
d & 10\,000 &  6\,000 & 78 & 90 & 35 & $5.5\times 10^{-1}$ \\
\hline
\end{tabular}
\label{param}
\end{table}

All the parameters for the three different wind regions are now fixed and summarized 
in Table \ref{param}. If this picture of the non-spherical wind is about correct, we
should be able to fit the luminosities of the remaining forbidden lines (i.e. from
S{\sc ii}, Cl{\sc ii}, Cl{\sc iii}, Ar{\sc iii}, and N{\sc ii}) with the same sets of 
parameters.  

The modeling of the [S{\sc ii}] lines was rather tricky, because they arise 
in two different regions: the hydrogen neutral disk and the outer parts of the 
(ionized) intermediate wind, but with no clear boundary conditions.
For the intermediate wind, S{\sc ii} can only arise if S{\sc iii} has recombined.
The saturation of the [S{\sc iii}] line luminosities happens at about 
$10^{6}$\,R$_{*}$ (Fig.\,\ref{sulfur3}) which is set as the minimum distance for 
the [S{\sc ii}] lines. For the emission from the disk the inner edge is set to the 
stellar surface. A reliable fit is found from the simultaneous fitting of both 
contributions where we found that the disk contributes less than half to 
the total emission and that S{\sc ii} has to recombine at a distance of about
6\,000\,R$_{*}$. The contribution from the intermediate wind which is more than half
of the total line luminosities comes from a very narrow (in radius) region, starting at 
a distance of about $1.05\times 10^{6}$\,R$_{*}$ just beyond the region of S{\sc iii} 
line luminosity saturation and extending to the edge of our slit which is at 
$1.13\times 10^{6}$\,R$_{*}$. Since the line luminosities from the intermediate wind 
did not saturate at the outer edge of our slit, we expect the [S{\sc ii}] lines to 
appear much more luminous when observed with a larger slit width. 

The predicted line luminosities of the nitrogen lines were much too high so that we 
had to reduce the N abundance to $0.5 \times $ solar to achieve good fits.

In Table\,\ref{lumis} we summarize all the observed (column 3) and 
modelled (column 4) line luminosities, as well as the ratio of observed over
modelled luminosity (column 5) and the location(s) of the ions in either the polar wind 
(p), intermediate wind (i) or disk (d) or a combination of two. We thereby distinguish 
two sets of lines: the lines which are used to derive the mass fluxes and temperature
distributions in the different wind regions are collected in the upper part, and the 
lines that are used for confirmation of the models and for derivation of the abundance 
(especially of N) are given in the lower part.





\begin{table}
\caption{Observed and modelled forbidden line luminosities. For the observed
values, we took the intensities given in Table\,\ref{Table_1} and assumed a 
distance of 2\,kpc. The fifth column contains the ratio of observed over 
modelled line luminosity and in the sixth column we describe in which part(s)
of the wind the line is produced: The classifications of p, i, and d mean 
polar wind, intermediate wind, and disk, respectively. We gathered in the upper
part the lines which are used to derive the mass fluxes and temperature distributions
in each wind zone while the lines in the lower part have been used to confirm the 
models and to derive (in the case of N) the abundance.}
\begin{tabular}{lccccc}
\hline
\hline
Ion & $\lambda$ (\AA)  &   $L_{\lambda}^{\rm obs}$  &  $L_{\lambda}^{\rm 
model}$ & ratio & region  \\
\hline
O{\sc iii}  & 4363 & $9.71\times 10^{33}$ & $2.90\times 10^{34}$ & 0.33 & p,i \\
O{\sc iii}  & 4959 & $3.58\times 10^{34}$ & $3.91\times 10^{34}$ & 0.92 & p,i \\
O{\sc iii}  & 5007 & $1.13\times 10^{35}$ & $1.13\times 10^{35}$ & 1.00 & p,i \\
O{\sc ii}   & 7319 & $7.23\times 10^{33}$ & $7.70\times 10^{33}$ & 0.94 & i \\
O{\sc ii}   & 7330 & $6.18\times 10^{33}$ & $6.17\times 10^{33}$ & 1.00 & i \\
O{\sc i}    & 5577 & $1.12\times 10^{32}$ & $2.33\times 10^{32}$ & 0.48 & d \\
O{\sc i}    & 6300 & $4.98\times 10^{32}$ & $4.86\times 10^{32}$ & 1.02 & d \\
O{\sc i}    & 6364 & $1.49\times 10^{32}$ & $1.60\times 10^{32}$ & 0.93 & d \\
S{\sc iii}  & 6312 & $2.95\times 10^{33}$ & $2.98\times 10^{33}$ & 0.99 & i \\
\hline
S{\sc ii}   & 4068 & $1.15\times 10^{33}$ & $4.57\times 10^{32}$ & 2.52 & i,d \\
S{\sc ii}   & 4076 & $3.43\times 10^{32}$ & $1.18\times 10^{32}$ & 2.90 & i,d \\
S{\sc ii}   & 6716 & $9.05\times 10^{31}$ & $1.05\times 10^{32}$ & 0.86 & i,d \\
S{\sc ii}   & 6731 & $2.13\times 10^{32}$ & $2.20\times 10^{32}$ & 0.97 & i,d \\
N{\sc ii}   & 5755 & $4.56\times 10^{33}$ & $4.57\times 10^{33}$ & 1.00 & i \\
N{\sc ii}   & 6548 & $4.84\times 10^{33}$ & $6.15\times 10^{33}$ & 0.79 & i \\
N{\sc ii}   & 6584 & $2.08\times 10^{34}$ & $1.81\times 10^{34}$ & 1.15 & i \\
Cl{\sc iii} & 5517 & $1.10\times 10^{32}$ & $4.23\times 10^{31}$ & 2.60 & p,i \\
Cl{\sc iii} & 5538 & $1.74\times 10^{32}$ & $1.74\times 10^{32}$ & 1.00 & p,i \\
Cl{\sc ii}  & 6153 & $3.88\times 10^{31}$ & $3.82\times 10^{31}$ & 1.01 & i \\
Ar{\sc iii} & 5193 & $1.16\times 10^{32}$ & $6.44\times 10^{32}$ & 0.18 & p,i \\
Ar{\sc iii} & 7136 & $7.37\times 10^{33}$ & $6.88\times 10^{33}$ & 1.07 & p,i \\
Ar{\sc iii} & 7753 & $1.49\times 10^{33}$ & $1.68\times 10^{33}$ & 0.89 & p,i \\
\hline
\end{tabular}
\label{lumis}
\end{table}

Inspecting Table\,\ref{lumis} shows that some of the modelled lines can be off by a 
factor of 2 or larger. Here, we want to give some arguments that might explain these 
differences.
                                                                                              
\begin{enumerate}
                                                                                              
\item We start with the [O{\sc i}] line at $\lambda = 5577$\,\AA~for which our
model predicts about twice as much luminosity than observed. This line corresponds
to the transition $5\longrightarrow 4$ in our adopted 5-level atom. There exists one
permitted transition between its upper level and a much higher level with wavelength
$\lambda = 1217.6$\,\AA~which falls into the wavelength range covered by a broadened
Ly\,$\alpha$ line ($\lambda_{\rm Ly\,\alpha} = 1215.6$\,\AA). The fifth level might
therefore be depopulated radiatively into this higher state from which several 
permitted lines arise, and the observable 5577\,\AA~line luminosity will decrease.
                                                                                              
\item An important source of error might be the foreground extinction used to de-redden
our data. We took the foreground extinction value from Costa et al. (\cite{Costa}) who
derived it from the Balmer lines which they assumed to arise in a spherically symmetric
constant density nebula. Since some of the modelled line luminosities arising in the 
blue part of the spectrum (at $\lambda < 5000$\,\AA)~are off compared to the observed 
luminosities, it seems that this might be a systematical error rather than a model 
error. However, our modeling procedure is not suitable to derive the real
extinction value. A slightly different extinction will certainly result in slightly
different model parameters, e.g. the mass fluxes in the different wind regions will
change individually. This makes it difficult to predict in which way the extinction
might be off, especially since some of the forbidden lines arise in two different
regions and have to be fitted simultaneously, as e.g. the [S{\sc ii}] lines.
Nevertheless, de-reddening with the correct extinction value is especially important 
if we have to compare lines 
from the same ion arising in the very blue part of the spectrum with lines coming from 
the very red part of the spectrum, which is the case e.g. for the lines of Ar{\sc iii} 
where we have a deviation of the blue line with respect to the red lines by a factor 
of 5! 
                                                                                              
\item An additional reason for the deviations might be the accuracy of the measured
line fluxes coming from different wavelength regions. As explained in 
Sect.\,\ref{observations}, the line fluxes from the high and low wavelength regions,
i.e. at wavelengths $\lambda < 5000$\,\AA~and $\lambda >
7000$\,\AA~have much larger uncertainties due to the lower S/N ratio and efficiency
of the CCD leading to larger errors in the measured line fluxes.
                                                                                              
\end{enumerate}
                                                                                              
Except of the [O{\sc i}] 5577\,\AA~line we think that both effects, a slightly different
foreground extinction and a larger uncertainty in the measured line fluxes, play a role
but it is difficult to disentangle the individual influence of these effects on a 
certain line.

From the results presented in Table\,\ref{param} for the different wind parts it 
is obvious that the mass flux varies with latitude
being lowest at the poles and highest at the equator with $F_{\rm m, disk}\simeq
8\,F_{\rm m, pole}$. The total mass loss rate of the star is given by the 
integral of the mass flux over the stellar surface which is
\begin{equation}
\dot{M} = 4\pi R_{*}^{2}\int\limits_{0}^{\pi/2}F_{\rm m}(\theta) \sin\theta d\theta
\end{equation}
Inserting our values given in Table\,\ref{param} we find a total mass loss rate
of the star of $\dot{M}\simeq (2.9\pm 0.6)\times 10^{-5}\,{\rm M}_{\odot}{\rm yr}^{-1}$.
The error results from the uncertainties in mass flux.

\subsection{The wings of H$\alpha$}\label{balmerwings}

With this complete wind scenario derived in the previous section we can return to 
the question whether the broad wings of the H$\alpha$ line are due to outflow or 
electron scattering. From our data we derive a contribution of about 10\%
of the broad wings to the total intensity which implies an electron scattering
optical depth of about 0.09 in and above the line formation region of H$\alpha$.
On the other hand, we can calculate the electron optical depth from
\begin{equation}
\begin{array}{l}
\tau_e=\int N_e(r) \sigma_e dr
\end{array}
\end{equation}
where $N_e(r)$ is the electron density distribution in the line of sight 
between the observer and the formation region of the line, and $\sigma_e$
is the electron scattering coefficient. Since we see Hen 2-90 more or less
edge-on, we can use the electron density distribution from the intermediate
wind and calculate backwards, i.e. from the observer into the wind, to determine
where in the wind most of the H$\alpha$ emission is produced. For our 
intermediate wind electron density distribution and the optical depth of 0.09
we find that the majority of the H$\alpha$ emission arises within 
about 4000\,R$_{*}$ around the star which is the region of highest
density and therefore the most plausible location. We can therefore state
that the wings of H$\alpha$ are indeed produced due to electron scattering
and not due to a high velocity wind. 

Additional confirmation of the broadening due to electron scattering rather than
due to high velocity outflow is provided by our wind scenario proposed above.
Our non-spherical wind model covers the complete surrounding of the star 
except of the highly-collimated jets which have been found to have real
outflow velocities of less than 400\,km\,s$^{-1}$ (see e.g. Sahai et al., 
\cite{Sahai02}). 
What we found from our wing velocity measurements in all the
different parts of the wind is: a terminal outflow velocity of 
about 85\,km\,s$^{-1}$ from the [O{\sc iii}] lines in the polar wind, of about 
40\,km\,s$^{-1}$ from the [N{\sc ii}] and [S{\sc iii}] lines in the 
intermediate wind, and about 35\,km\,s$^{-1}$ from the [S{\sc ii}] and 
[O{\sc i}] lines in the disk. We could not identify any additional wind 
component which might have an outflow velocity as high as 1800\,km\,s$^{-1}$.


\section{Discussion}

\subsection{The non-spherical wind scenario for Hen 2-90}\label{validity}

For the model calculations we use the geometry and ionization structure
based on the HST image (Fig.\,\ref{HST}) which clearly shows a polar wind, an 
intermediate wind
and disk so that the geometry used seems to be well justified.

Such a non-spherical wind model with latitude dependent temperature, terminal
wind velocity, and mass flux might be understood in terms of a rotating star.
According to the van Zeipel theorem, rapid rotation will lead to polar brightening.
This means that the surface temperature of the star can be much hotter on the 
pole compared to the equator. Our calculations show that in equatorial regions
the surface temperature should not be higher than about 10\,000\,K to guarantee
that hydrogen is neutral. In polar directions we found that the 
terminal temperature should not be lower than about 10\,000\,K which means that
the surface temperature can be much hotter. We do not have a good handle on the 
real polar surface temperature because the forbidden emission lines are produced far 
away from the star where the temperature dropped already to its terminal value. 
In our calculations we used the effective temperature of 50\,000\,K
found by Kaler \& Jacoby (\cite{Kaler}) with the Zanstra method. Cidale et al. 
(\cite{Cidale}) derived a much lower effective temperature of only 32\,000\,K using 
the energy balance method. We tested our model for different polar surface
temperatures in the range $T_{0} = 15\,000 - 50\,000$\,K. We found, that for 
different surface temperatures large differences occur in the line luminosities 
close to the star. However, in the regions of our interest, where the lines
saturate, the line luminosities are all about equal. This implies that from our
observations we cannot draw any conclusion on the real polar surface temperature.

Rotation not only influences the surface temperature structure, but also the 
terminal wind velocities resulting in a decrease in terminal velocity from pole to 
equator (see e.g. Lamers \& Cassinelli, \cite{Lamers99}; Kraus \& Lamers, \cite{Kraus05}). 
This is exactly what we found from our observations, when we take into 
account the known inclination of the system as suggested from the HST image. 
 
The ratio of polar (85\,km\,s$^{-1}$) to disk terminal velocity (35\,km\,s$^{-1}$)
of about 2.4 derived from the forbidden emission lines is consistent with a 
rotation speed of about 75 -- 80\,\% of the critical velocity (Kraus \& Lamers, 
\cite{Kraus05}). Such high velocities are observed e.g. for the classical Be stars 
(see e.g. Porter \& Rivinius, \cite{Porter}). Some of those stars
have recently been suggested to rotate even close to break-up, i.e. with 
almost critical velocity (Townsend et al., \cite{Townsend}).

An additional effect of rotation is that it leads to a latitude dependent mass flux 
which is also the case for Hen 2-90. A mechanism like the wind compressed disk 
(Bjorkman \& Cassinelli, \cite{Bjorkman}) or the rotation induced bistability  
(Lamers \& Pauldrach, \cite{Lamers91}), or a combination of both can result in the 
observed high density equatorial outflowing disk.

The combination of the latitude dependent mass flux and surface temperature also 
explains why we see different ionization structures at different latitudes. The 
lower surface temperature results in less ionizing UV photons, and the higher mass 
flux into a higher shielding density which in summary reduces the degree of ionization.

For our model calculations the outflowing disk is assumed to be neutral in hydrogen.
Kraus \& Lamers (\cite{Kraus03}) showed that in the case of supergiants 
the disk created by an equatorial wind can indeed become neutral just above 
the stellar surface if the equatorial mass flux is high enough to prevent ionizing 
stellar photons from penetrating into the high density equatorial disk wind.
Although Hen 2-90 is hotter than a typical supergiant (but its effective temperature
might well be only 32\,000\,K instead of 50\,000\,K as stated above), it has also
a much smaller radius leading to an enormous equatorial mass flux whose material
is indeed dense enough to shield the disk material efficiently from stellar irradiation.
Consequently, the disk will stay neutral even close to the stellar surface.

The terminal temperature found for the gas disk is about 6\,000\,K. From our modeling  
we found that the luminosities of the forbidden lines coming from the disk saturate at 
a distance of about $2\times 10^{4}$\,R$_{*}$. The total size of the disk is, however, 
much larger, at least $5\times 10^{5}$\,R$_{*}$. The ISO spectrum of Hen 2-90 shows a 
strong IR excess due to hot circumstellar dust which can only be located in the disk. 
The evaporation temperature of dust is between 1\,500\,K and 2\,000\,K, depending on its 
composition. In the temperature region between the atomic and the dust dominated regions, 
the atoms might be locked into molecules that can form (and survive) at temperatures 
below about 5\,000\,K. Our results are therefore also consistent with a circumstellar
dust disk where the dust forms far away from the star, at distances $r_{\rm dust} \gg
2\times 10^{4}$\,R$_{*} \simeq 35$\,AU. A detailed study of the ISO spectrum 
to calculate the infrared emission coming from the predicted dust disk is necessary 
and currently under investigation (Borges Fernandes \& Lorenz-Martins, in preparation).

It would be interesting to see whether the ionization structure we found from our 
modeling of the forbidden emission lines might be reproduced by more sophisticated 
ionization structure calculations in such a non-spherically
symmetric wind scenario. Unfortunately, as far as we know,
there exist no numerical codes to perform the necessary 3D radiation transfer needed
for non-spherical ionization structure calculations.

Finally, we qualitatively discuss how the variety of the line profiles, shown in 
Sect.\,\ref{atlas}, might be produced in our non-spherical wind scenario. Even 
though the central star might be rapidly rotating, we do not think that (especially
the double peaked) line profiles mirror the stellar rotation or are due to rotation 
at all (although we cannot ad hoc exclude a rotational contribution) because the effective 
temperature of about 50\,000\,K indicates, that the star must have a line-driven wind. 
The acceleration of the wind to its terminal velocity happens within only a few
stellar radii, leading to a radially outflowing wind.

{\bf Double-peaked lines:} 
these lines are mainly produced 
in the intermediate wind (some of them with a gaussian contribution 
from the polar wind) which can be regarded as a thick expanding torus. 
Since we see this torus edge-on, the emission lines formed within it
show double-peaked profiles. Since the maximum (terminal) wind velocity 
of the intermediate wind is about 45\,km\,s$^{-1}$, the peak separation 
will be of order 40 to 50\,km\,s$^{-1}$, what is observed.

{\bf Broad single-peaked lines:} 
they are found for the hydrogen Balmer lines. As discussed
already in Sect.\,\ref{balmerwings}, the broad wings of H$\alpha$ are due
to electron scattering with a scattered intensity of about 10\%.
With increasing quantum number, the Balmer line intensities decrease and
a 10\% contribution in the broad wings becomes much harder to detect above
the continuum. That's why with increasing quantum number the derived wing
velocities of the Balmer lines decrease until they adapt to the
terminal wind velocities derived from the forbidden lines (see e.g. line H8
in Table\,\ref{Table_2}). 

{\bf Narrow single-peaked lines:} these lines are mainly formed in the disk
(even though some of them can have small contributions from the intermediate wind).
The lines formed in the outflowing edge-on seen disk should also show 
double-peaks as the lines formed in the intermediate wind. However, since
the disk outflow velocity is smaller than the velocity in the intermediate
wind, the two peaks merge leading to a single-peaked line profile.

The variety of profiles observed can be explained with our model assumptions.
However, we have no explanation for the asymmetric line profiles with a strong
red peak. Probably, our scenario is not completely correct, and effects like
clumping in the wind might be worth being investigated.
 
A quantitative analysis of the line profiles would be necessary, but this is 
with our data difficult and beyond the scope
of this paper, because the profiles are derived from the FEROS spectrum which
is not flux calibrated, while the line luminosities have been derived from the
low-resolution Cassegrain spectrum, which makes it impossible to model
line luminosities and profiles simultaneously.

\subsection{Hen 2-90: a symbiotic object or a compact planetary nebula ?}\label{nature}

In the literature Hen 2-90 has been classified either as a (compact) planetary
nebula (e.g. Costa et al., \cite{Costa}; Maciel, \cite{Maciel}; Lamers et al., 
\cite{Lamers98}) or as a symbiotic object, i.e. a binary consisting of a
cool giant and a hot component with an accretion disk (Sahai \& Nyman, \cite{Sahai00};
Guerrero et al., \cite{Guerrero}). 

The HST image (Fig.\,\ref{HST}, Sahai et al., \cite{Sahai02}) clearly shows the 
presence of a nebula bisected by a disk, with a highly collimated and bipolar jet 
with several pairs of knots at both sides. This scenario could be explained assuming 
that Hen 2-90 is either a compact planetary nebula, where the wind asymmetries started 
during the AGB phase, or a binary system, where the jets are caused by an accretion disk.

During our modelling procedure we have found, as also cited by Costa et al. (\cite{Costa}) 
and Maciel (\cite{Maciel}), a depletion of N and O by a factor of 2 and 3.3, 
respectively, with respect to the electron density with S, Ar and Cl having 
normal (i.e. solar) abundances. In addition, the FEROS spectrum contains only one 
single permitted carbon line, C{\sc ii} $\lambda$6578. This line has previously been 
detected by Costa et al. (\cite{Costa}) and Guerrero et al. (\cite{Guerrero}), but it is 
very weak in the FEROS spectrum (just at the detection limit) and absent
in the Cassegrain spectrum. We can only estimate its line flux and we found 
about 0.06 for the observed flux, and about 0.02 for the dereddened flux in the scaling 
of Table\,\ref{Table_1} (H$\beta = 100$). The uncertainty of these line fluxes is 
about 50\%. No forbidden emission line of carbon was detected, even though
in the optical spectrum we would expect to observe several [C{\sc i}] lines coming
from the disk. We applied our disk wind model to calculate the 
line luminosities for these carbon lines, and found that these lines would only show up 
in the spectrum for a carbon abundance of at least twice solar.
Inspection of the ISO spectrum of Hen 2-90 reveales emission from oxygen-rich
dust (Borges Fernandes \& Lorenz-Martins, in preparation). Oxygen-rich dust can 
only be the dominant dust component in case ${\rm O/C} > 1$ during the ejection time 
of the dust forming material. We found an underabundance of O by more 
than a factor of 3. Together with the fact that the dust composition should 
mirror the actual wind composition (unless the star has changed its 
composition from O-rich to C-rich since the ejection of the dust forming 
material, which is rather unlikely), 
we conclude that (i) the progenitor star that ejected the dust forming material
cannot have been carbon rich, and (ii) the actual carbon 
abundance must be less than 0.6 solar to guarantee ${\rm O/C} > 1$.
Assuming a standard scenario with dredge-ups, we can understand the depletion 
of C and O, however, we have no explanation for the fact that N is also 
depleted. 
There are some {\it B-type} post-AGB stars in the halo of our Galaxy showing the same 
behaviour (Moehler \& Heber, \cite{Moehler}), and recently Lennon et al. (\cite{Lennon})
reported on a Be star with an unexpected low N abundance. However, the origin of those 
anomalies is still poorly understood. Rotation or a possible binarity could play 
an important role. Unfortunately, as far as we know, there exist no stellar evolution
models that follow the whole life sequence of rapidly rotating intermediate mass stars,
for which Hen 2-90 seems to be a candidate. 

Another point that should be commented is that from our spectra, that cover the 
complete optical wavelenght range, we have identified no lines from ions with ionization 
potential higher than $\sim 40$\,eV and consequently no He\,{\sc ii} lines. In addition,
there is no indication for the presence of TiO bands either. These two characteristics
are the main signature of a symbiotic system. Since we see the disk of Hen 2-90 
almost edge-on it might be possible that the disk hides the atmosphere of a cool giant, 
where the TiO bands are formed, and also absorb the He\,{\sc ii} emission from a hot 
component. However, the fact that we don't see hints for any other line from ions
with ionization potential higher than $\sim 40$\,eV leads us to the conclusion that 
these ions (and therefore also He\,{\sc ii}) do not exist in the wind of Hen 2-90 
indicating that (i) the effective temperature of Hen 2-90 is indeed much lower than 
the 50\,000\,K found from the Zanstra method and may well be of order 32\,000\,K as
derived by Cidale et al. (\cite{Cidale}) from the energy balance, and (ii) the 
symbiotic system is not the favoured classification. 

Besides the absence of clear indications for a classification as a symbiotic object,
our results favour the conclusion of Hen 2-90 being a compact planetary nebula and we
want to mention two more features that support this idea:
                                                                                            
1. The mass loss rate of $\sim 3\times 10^{-5}$M$_{\odot}$yr$^{-1}$ found for Hen 2-90
coincides with those found for proto-planetary nebulae with axially symmetric dust shells
resulting from a so-called superwind (see e.g. Meixner et al.,\,\cite{Meixner}).
                                                                                            
2. An observational curiosity in the spectrum of Hen 2-90 is the fact that the
[O{\sc iii}] 5007\,\AA~line is less intense than the H$\alpha$ line, although for
a compact planetary nebula it is usually the other way round. However, in
our forbidden line analysis we found that the abundance of O had to be reduced
to only 0.3 solar to achieve reasonable line luminosity fits. This means, that if
Hen 2-90 had a \lq normal\rq~O abundance, the 5007\,\AA~line would be about 3.3
times stronger than observed, and consequently be higher than the H$\alpha$ flux.
The anomalous line flux is therefore due to the underabundance of O.

We cannot definitely exclude a binary nature of Hen 2-90 (see also the discussion
in Kraus et al., \cite{Krausetal05}). Unfortunately,
the optical continuum of Hen 2-90 is completely flat, making it impossible to 
disentangle individual components. In addition, the optical continuum is mainly 
due to free-free and free-bound emission in the optically thick wind, rather
than by the stellar continuum.

Observed features, that might hint to a binary nature, are the jet-like structure 
with blob ejections (Sahai, \cite{Sahai}; Sahai et al., \cite{Sahai02}),
and the DENIS NIR photometry cited by Guerrero et al. (\cite{Guerrero}):

{\bf Jet-like structure and blob ejection:}
The jet structure seen in Hen 2-90 is very collimated, and the blobs are found to
be ejected regularly on a 40 year timescale. This perfect alignment of the jet
axis and the blobs is unique. All PNe which are known to be binaries
show either asymmetric or point-symmetric jet and blob ejection.
In addition, to date no X-ray emission coming from an accretion disk has been 
detected from the system which might indicate a completely different formation mechanism.
Garc$\acute{\rm \i}$a-Segura et al. (\cite{Garcia}) modeled the jet and knots of Hen 2-90
using the concept of magnetohydrodynamics in the wind of a single star and under
the assumption of a solar-like magnetic cycle. This model does not need a binary 
for the jet production, because the jet formation is not driven by an accretion disk.
A binary component might be helpful in the first instance, to spin-up the star to high 
rotation rates (e.g. in a merger process) and, therefore, to increase the stellar 
magnetic field to the values needed for their model calculations. Such a spin-up might 
also explain the high rotation velocity of the central star which we claimed in the 
previous section and derived from the observed terminal velocities. 

{\bf DENIS NIR photometry:}
Schmeja \& Kimeswenger (\cite{Schmeja}) used the new DENIS survey data to probe 
different types of PNe. Their Fig.\,1 shows a distinction between genuine PNe,
symbiotic Miras and IR-[WC] PNe on the basis of a colour-colour plot. With the 
published DENIS photometry data of Hen 2-90 (Schmeja \& Kimeswenger, \cite{Schmeja2}),
and corrected for the extinction value of Costa et al. (\cite{Costa}), Hen 2-90
falls into the intermediate region between the genuine PNe and the IR-[WC] PNe.
However, the IR-[WC] PNe are supposed to have strong PAH emission which increases
their K-band flux and shifts them off the genuine PNe region. A pollution
of the K band emission might also be present for Hen 2-90 which shows a huge 
zoo of emission features in the IR. Subtraction of this polluting emission might 
shift Hen 2-90 back into the regime of the genuine PNe. In addition, the carbon
abundance found for Hen 2-90 is way too small to classify it as a carbon rich
star.

From our observations, modeling and the above discussion we conclude that 
Hen 2-90 must be at least an evolved and probably rapidly rotating object, and the 
classification as a compact planetary nebula seems to be favourable. The question 
whether this compact planetary nebula is part of a binary system can only be solved 
if clear direct indications of a companion star are observed, which is up to now 
not the case but makes it worth looking at this object in much more detail.

\section{Conclusions}

In this paper we studied the non-spherical mass loss history of Hen 2-90 by
splitting the analysis into a qualitative and a quantitative part. 

In the first part,
we presented high and low resolution optical observations of the innermost 
non-spherical wind structure around Hen 2-90. The spectra contain a huge number of 
forbidden and permitted emission lines of different ionization states justifying
the classification of Hen 2-90 as an object showing the B[e] phenomenon. The emission 
lines can be separated into four groups according to the different shape of their 
profiles: broad and narrow single-peaked lines, double-peaked lines, and lines with 
a well pronounced shoulder. This variety of line profiles indicates a complex 
structure of the circumstellar medium. There are no absorption lines present in the 
spectrum.

In the second part of the paper we performed a detailed analysis of the forbidden 
emission lines. The wind geometry used is based on the HST image which reveals a 
non-spherical wind structure consisting of a polar wind, an outflowing disk and an 
intermediate wind in between these two (see Fig.\,\ref{HST}). In all three parts, the 
ionization structure is different, 
indicating a latitude dependence of the surface temperature and the mass flux. 
These assumptions could be confirmed by our forbidden line analysis and might be 
interpreted in terms of a rapidly rotating (75 -- 80\% critical) underlying star. 
We could determine  
mass fluxes of $0.07\pm 0.01$\,g\,s$^{-1}$cm$^{-2}$, $0.15\pm 0.03$\,g\,s$^{-1}$cm$^{-2}$, 
and $0.55\pm 0.11$\,g\,s$^{-1}$cm$^{-2}$ (i.e. 
$7.99\times 10^{-7}$\,M$_{\odot}$yr$^{-1}$steradian$^{-1}$,  
$1.66\times 10^{-6}$\,M$_{\odot}$yr$^{-1}$steradian$^{-1}$, and 
$6.10\times 10^{-6}$\,M$_{\odot}$yr$^{-1}$steradian$^{-1}$) 
for the polar, intermediate and disk wind, 
respectively. The surface temperature might change from 50\,000\,K -- 32\,000\,K (or less) 
at the pole to about 10\,000\,K at the equator, and the terminal velocities are 
85\,km\,s$^{-1}$ (polar wind), 45\,km\,s$^{-1}$ (intermediate wind) and 35\,km\,s$^{-1}$ 
(disk). The total mass loss rate is found to be $(2.9\pm 0.6)\times 
10^{-5}$\,M$_{\odot}$yr$^{-1}$.

From the almost absence of observable carbon lines in our spectra and from the fact
that Hen 2-90 shows clearly O-rich dust in the ISO spectrum, we could restrict the 
C abundance to a value less than about 0.6 solar.
In addition we found that the O abundance is 0.3 solar and the N abundance is about 0.5
solar leading to an enhanced N/O ratio of 5/3 with respect to 
the solar value, while Ar, S and Cl could be modelled with solar abundances.
The depletion of C and O follow from stellar evolution, 
but for the depletion of N we have no satisfying explanation; it cannot be explained 
by a standard stellar evolution scenario with dredge-ups. 

From our observations and modeling
results we can conclude that Hen 2-90 must be an evolved object with most probably a
rapidly rotating central star and we favour the interpretation of Hen 2-90 being a compact 
planetary nebula. Whether it is part of a binary system is still an unsolved 
question. More observations and better evolutionary models for rapidly rotating stars are 
certainly needed for a better comprehension of this really curious object.

%

\begin{acknowledgements}
      We would like to thank the unknown referee for critical remarks and suggestions
      that
      have led to a significant improvement of this paper. M.K. also thanks 
      Guillermo Garc$\acute{\rm \i}$a-Segura for helpful discussions on PNe.
      M.K. was supported by the German \emph{Deut\-sche For\-schungs\-ge\-mein\-schaft, 
      DFG\/} grant number Kr~2163/2--1 and by the Nederlandse Organisatie voor
      Wetenschappelijk Onderzoek (NWO) grant No.\,614.000.310.
      M.B.F. acknowledges financial support from
      \emph{CAPES} (Ph.D. studentship). M.B.F. also acknowledges Utrecht University  
      for the warm hospitality during his one year stay there.
\end{acknowledgements}


\begin{table}
\caption{Emission line intensities relative to H$\beta$ = 100 obtained
with B\&C spectrograph.}
\begin{tabular}{cccc}
\hline
\hline
Wavelength (\AA)  &   I$_{\rm obs}$($\lambda$)  & I$_{\rm corr}$($\lambda$)  & Identification \\
\hline
3833.8  &  3.14 & 12.88   &   H9 3835.4 \\
3867.1  &  1.67 &  6.37   &   He\,{\sc i} (m20) 3867.5 \\
3887.5  &  7.20 & 26.22   &   He\,{\sc i} (m2) 3888.7 \\
        &       &   &   H8 3889.1 \\
3968.6  &  6.06 & 20.37   &   H$\epsilon$ 3970.1 \\
4025.5  &  1.15 &  3.48   &   He\,{\sc i} (m18) 4026.2 \\
4068.0  &  0.95 &  1.78   &   S\,{\sc ii} [m1F] 4068.6 \\
4075.8  &  0.33 &  0.53   &   S\,{\sc ii} [m1F] 4076.2 \\
4101.4  &  11.24 & 33.26   &   H$\delta$ 4101.7 \\
4120.5  &  0.19 &  0.45   &   O\,{\sc ii} (m20) 4119.2 \\
        &       &   &   O\,{\sc ii} (m20) 4120.3 \\
        &       &   &   O\,{\sc ii} (m20) 4120.6 \\
        &       &   &   He\,{\sc i} (m16) 4121.0 \\
        &       &   &   S\,{\sc ii} (m2) 4121.0 \\
4143.8  &  0.30 & 0.85   &   He\,{\sc i} (m53) 4143.8 \\
4178.8  &  0.23 &  0.41   &   Fe\,{\sc ii} (m28) 4178.9\\
        &    &        &   Fe\,{\sc ii} [m21F] 4177.2 \\
        &    &      &   Fe\,{\sc ii} [m23F] 4179.0 \\
4233.4  &  0.12 &  0.17   &   Fe\,{\sc ii} (m27) 4233.2 \\
4245.0  &  0.14 &  0.26   &   Fe\,{\sc ii} [m21F] 4244.8 \\
4288.0  &  0.20 &  0.38   &   Fe\,{\sc ii} [m7F] 4287.4 \\
4341.3  &  24.40 & 54.89   &   H$\gamma$ 4340.5 \\
4364.0  &  4.12 & 15.03   &   O\,{\sc iii} [m2F] 4363.2 \\
4388.6  &  0.51 &  0.99   &   He\,{\sc i} (m51) 4387.9 \\
4416.3  &  0.46 &  0.83   &   Fe\,{\sc ii} (m27) 4416.8 \\
        &       &   &   Fe\,{\sc ii} [m6F] 4414.5 \\
        &       &   &   Fe\,{\sc ii} [m6F] 4416.3 \\
4472.7  &  3.68 &  6.68   &   He\,{\sc i} (m14) 4471.7 \\
4490.9  &  0.08 &  0.21   &   Fe\,{\sc ii} [m6F] 4488.8 \\
        &       &   &   Fe\,{\sc ii} (m37) 4491.4 \\
4520.8  &  0.26 &  0.36   &   Fe\,{\sc ii} (m37) 4520.2 \\
        &       &   &   Fe\,{\sc ii} (m38) 4522.6 \\
4554.3  &  0.23 &  0.38  &   S\,{\sc iii} (m2) 4552.7 \\
4584.7  &  0.22 &  0.33   &   Fe\,{\sc ii} (m37) 4582.8 \\
        &       &   &   Fe\,{\sc ii} (m38) 4583.8 \\
        &       &   &   Fe\,{\sc ii} (m26) 4584.0 \\
4596.3  &  0.16 & 0.22   &   Fe\,{\sc ii} (m37) 4595.7 \\
4608.1  &  0.23 & 0.37   &   N\,{\sc ii} (m5) 4607.2 \\
4630.8  &  0.17 & 0.27    &   Fe\,{\sc ii} (m37) 4629.4 \\
        &       &   &   N\,{\sc ii} (m5) 4630.5 \\
4641.7  &  0.11 & 0.20    &   N\,{\sc iii} (m2) 4640.6 \\
        &       &   &   O\,{\sc ii} (m1) 4641.8 \\
        &       &   &   N\,{\sc iii} (m2) 4641.9 \\
\hline
\end{tabular}
\end{table}
                                                                                
\setcounter{table}{3}
                                                                                
\begin{table}[h!]
\caption{({\it continued})}
\vspace{-0.4cm}
\begin{tabular}{cccc}
\hline
\hline
Wavelength (\AA)  &   I$_{\rm obs}$($\lambda$)  &  I$_{\rm corr}$($\lambda$)  &  Identification \\
\hline
4659.3  &  3.62 & 4.31    &   Fe\,{\sc iii} [m3F] 4658.1 \\
4702.4  &  1.51 & 1.92    &   Fe\,{\sc iii} [m3F] 4701.5 \\
4713.5  &  0.81 & 1.05    &   He\,{\sc i} (m12)  4713.4 \\
4734.4  &  0.71 & 0.86    &   Fe\,{\sc ii} (m43) 4731.4 \\
4755.4  &  0.59 & 0.70    &   Fe\,{\sc iii} [m3F] 4754.7 \\
4770.6  &  0.62 & 0.80    &   Fe\,{\sc iii} [m3F] 4769.4 \\
4778.3  &  0.41 & 0.51    &   S\,{\sc ii} (m8) 4779.1 \\
        &       &   &   N\,{\sc ii} (m20) 4779.7 \\
4815.3  &  0.09 & 0.09    &   Fe\,{\sc ii} [m20F] 4814.6 \\
4861.7  & 100.00 & 100.00   &   H$\beta$ 4861.3 \\
4881.3  &   0.16 & 0.19   &   Fe\,{\sc iii} [m2F] 4881.0 \\
4904.8  &   0.20 & 0.23    &   Fe\,{\sc ii} [m20F] 4905.4 \\
4922.5  &   1.72 & 1.97    &   He\,{\sc i} (m48) 4921.9  \\
        &        &  &   Fe\,{\sc ii} (m42) 4923.9 \\
4930.5  &   0.42 & 0.43   &   Fe\,{\sc iii} [m1F] 4930.5 \\
4959.0  &  63.49 & 55.37   &   O\,{\sc iii} [m1F] 4958.9 \\
5001.1  & 0.32  & 0.26   &   N\,{\sc ii} (m19) 5001.1 \\
5006.8  & 182.80  & 149.08   &   O\,{\sc iii} [m1F] 5006.9 \\
5011.3  &  2.18 &  1.78   &   Fe\,{\sc iii} [m1F] 5011.3 \\
5015.7  &  4.22 &  3.45   &   He\,{\sc i} (m4) 5015.7 \\
5018.4  &  0.28 &  0.23   &   Fe\,{\sc ii} (m42) 5018.4 \\
5041.4  &  0.60 &  0.43   &   Si\,{\sc ii} (m5) 5041.1 \\
5047.6  &  0.30 &  0.17   &   S\,{\sc ii} (m15) 5047.3 \\
5055.8  &   0.56 & 0.30   &   Si\,{\sc ii} (m5) 5056.4 \\
5084.4  &   0.21 & 0.19   &   Fe\,{\sc iii} [m1F] 5084.8 \\
5158.9  &   0.31 & 0.21   &   Fe\,{\sc ii} [m18F] 5158.0 \\
5168.1  &   0.26 & 0.18   &   Fe\,{\sc ii} (m42) 5169.0  \\
5191.6  &   0.35 & 0.18   &   Ar\,{\sc iii} [m1F] 5193.3 \\
5197.9  &   0.57 & 0.28    &   Fe\,{\sc ii} (m49) 5197.6 \\
5233.9  &   0.36 & 0.26   &   Fe\,{\sc ii} (m49) 5234.6 \\
5270.2  &   3.94 & 2.93   &   Fe\,{\sc iii} [m1F] 5270.4 \\
5316.3  &   0.73 & 0.40   &   Fe\,{\sc ii} (m49) 5316.6 \\
5332.7  &   0.27 & 0.10   &   Fe\,{\sc ii} [m19] 5333.7 \\
5363.0  &   0.29 & 0.15   &   Fe\,{\sc ii} (m48) 5362.9 \\
        &        &  &   Fe\,{\sc ii} [m17F] 5362.1 \\
5375.1  &   0.13 & 0.06   &   Fe\,{\sc ii} [m19F] 5376.5 \\
5411.5  &   0.41 & 0.21   &   Fe\,{\sc iii} [m1F] 5412.0 \\
5425.4  &   0.12 & 0.05  &   Fe\,{\sc ii} (m49) 5425.3 \\
5432.2  &   0.12 & 0.04   &   Fe\,{\sc ii} (m55) 5432.9 \\
        &        &  &   Fe\,{\sc ii} [m18F] 5433.2 \\
5475.6  &   0.26 & 0.05   &   Fe\,{\sc ii} (m49) 5477.7  \\
5505.4  &   0.26 & 0.06   &   Cr\,{\sc iii} [m2F] 5505.1 \\
5517.6  &   0.56 & 0.17   &   Cl\,{\sc iii} [m1F] 5517.2 \\
5535.5  &   0.71 & 0.27   &   Cl\,{\sc iii} [m1F] 5537.7 \\
\hline
\end{tabular}
\end{table}
                                                                                
\setcounter{table}{3}
                                                                                
\begin{table}[h!]
\caption{({\it continued})}
\vspace{-0.4cm}
\begin{tabular}{cccc}
\hline
\hline
Wavelength (\AA)  &   I$_{\rm obs}$($\lambda$)  &  I$_{\rm corr}$($\lambda$)  &  Identification \\
\hline
5551.3  &   0.43 & 0.10   &   N\,{\sc ii} (m63) 5552.5 \\
5577.7  &   0.08 & 0.02   &   O\,{\sc i} [m3F] 5577.4 \\
5665.9  &   0.20 & 0.08   &   N\,{\sc ii} (m3) 5666.4 \\
5677.8  &   0.39 & 0.16   &   N\,{\sc ii} (m3) 5676.0 \\
5713.5  &   0.36 & 0.11    &   Fe\,{\sc ii} [m2F] 5713.4 \\
5754.5  &   12.16 & 7.06    &   N\,{\sc ii} [m3F] 5754.8  \\
5875.2  &  30.46 & 13.93    &   He\,{\sc i} (m11) 5875.6  \\
5956.8  &   0.47 & 0.21    &   Si\,{\sc ii} (m4) 5957.6 \\
5978.8  &   0.79 & 0.33    &   Si\,{\sc ii} (m4) 5979.0 \\
5999.8  &   0.35 & 0.12    &   Ni\,{\sc iii} 6000.2 \\
6095.8  &   0.12 & 0.03    &   S\,{\sc ii} (m13) 6097.1 \\
6124.5  &   0.23 & 0.06    &   S\,{\sc ii} (m13) 6123.4 \\
6152.5  &   0.26 & 0.06    &   Cl\,{\sc ii} [m3F] 6152.9 \\
6248.4  &   0.29 & 0.07    &   Fe\,{\sc ii} (m74) 6247.6 \\
6300.4  &   1.93 & 0.77   &   O\,{\sc i} [m1F] 6300.2 \\
6312.0  &   12.84 & 4.57   &   S\,{\sc iii} [m3F] 6311.9 \\
6347.1  &   1.06 & 0.35   &   Si\,{\sc ii} (m2) 6347.1 \\
6364.0  &   0.67 & 0.23   &   O\,{\sc i} [m1F] 6363.9 \\
6371.4  &   0.72 & 0.25    &   Fe\,{\sc ii} (m40) 6369.5 \\
        &        &  &   Si\,{\sc ii} (m2) 6371.4 \\
6384.6  &   0.55 & 0.16   &   Fe\,{\sc ii} 6383.8 \\
6401.4  &   0.26 & 0.07   &   Ni\,{\sc iii} 6401.5 \\
6456.6 &   0.30 & 0.08    &    Fe\,{\sc ii} (m74) 6456.4 \\
6483.2 &   0.17 & 0.05    &    Uid \\
6492.3 &   0.31 & 0.09    &    Fe\,{\sc ii} 6493.1 \\
6516.9 &   0.12 & 0.03    &    Fe\,{\sc ii} (m40) 6516.1 \\
6533.3 &    0.25 & 0.08   &    Ni\,{\sc iii} 6533.9 \\
6548.1  &   26.90 & 7.49   &   N\,{\sc ii} [m1F] 6548.1  \\
6563.0  &  1200.00 & 317.53   &   H$\alpha$ 6562.8 \\
6583.6  &  120.00 & 32.15   &   N\,{\sc ii} [m1F] 6583.6  \\
6678.0  &  10.29 & 3.08    &   He\,{\sc i} (m46) 6678.2  \\
6716.2  &  0.51 & 0.14    &   S\,{\sc ii} [m2F] 6716.4  \\
6730.5  &  1.25 &  0.33   &   S\,{\sc ii} [m2F] 6730.8 \\
6746.9  &  0.15 &  0.04   &   Cr\,{\sc iv} [m2F] 6746.2 \\
6793.3  &  0.13  & 0.04   &   Fe\,{\sc i} 6793.3 \\
6826.9  &  0.15  & 0.03   &   Fe\,{\sc iii} 6826.2 \\
6895.2  &  0.14 &  0.04   &   Fe\,{\sc ii} [m14F] 6896.2 \\
6914.8  &  0.57 & 0.14     &   Cr\,{\sc iv} [m2F] 6915.6 \\
6944.9  &  0.15 &  0.03   &   Fe\,{\sc ii} [m43F] 6944.9 \\
6996.2  &  0.31 &  0.07   &   [Fe\,{\sc iv}] 6997.1 \\
7001.9   &  0.14 &  0.03   &   O\,{\sc i} (m21) 7001.9   \\
7064.0  &  16.03 & 3.74    &   He\,{\sc i} (m10) 7065.2   \\
7087.3  &  0.11 & 0.03    &   Cr\,{\sc iv} [m2F] 7087.1 \\
7109.7  &  0.19 & 0.05   &   Uid \\
\hline
\end{tabular}
\end{table}
                                                                                
\setcounter{table}{3}
                                                                                
\begin{table}[h!]
\caption{({\it continued})}
\vspace{-0.4cm}
\begin{tabular}{cccc}
\hline
\hline
Wavelength (\AA)  &   I$_{\rm obs}$($\lambda$)  &  I$_{\rm corr}$($\lambda$)  &  Identification \\
\hline
7134.3  &  50.46 & 11.41   &   Ar\,{\sc iii} [m1F] 7135.8 \\
7154.2  &  0.32 &  0.07   &   Fe\,{\sc ii} [m14F] 7155.1 \\
7169.2  &  0.21 & 0.05   &   Fe\,{\sc ii} [m14F] 7171.9 \\
7181.8  &  0.10 &  0.02   &   Fe\,{\sc ii} (m72) 7181.2 \\
7232.9  &  0.74 &  0.14   &   Cr\,{\sc iv} [m1F] 7233.4 \\
7253.1  &  0.39 &  0.07   &   O\,{\sc i} (m20) 7254.2 \\
7279.7  &  2.03 & 0.41    &   He\,{\sc i} (m45) 7281.4  \\
7296.2  &  0.16 &  0.03    &   Uid \\
7317.9  &  57.34 & 11.19   &   O\,{\sc ii} [m2F] 7318.6 \\
7328.2  &  48.95 & 9.56   &   O\,{\sc ii} [m2F] 7329.9 \\
7375.8  &  0.31 &  0.06   &   Fe\,{\sc ii} 7376.5 \\
7388.5  &  0.42 &  0.08   &   Fe\,{\sc ii} [m14F] 7388.2 \\
7409.6  &  0.15 & 0.03 & Si\,{\sc i} (m23) 7409.1 \\
7422.6  &  0.13  & 0.02 & [Fe\,{\sc iii}] 7422.6 \\
7431.2  & 0.27  & 0.04 & Fe\,{\sc ii} 7431.35 \\
7449.2  & 0.22  & 0.02 & Fe\,{\sc ii} (m73) 7449.3 \\
        &       &      & Fe\,{\sc ii} [m30F] 7449.6 \\
7465.8  & 0.23  & 0.03 & Ni\,{\sc i} [m2F] 7464.4 \\
7478.1  & 0.13  & 0.02 & Fe\,{\sc i} 7478.9 \\
7498.0  &  0.54 &  0.08   &   Fe\,{\sc ii} [m3F] 7497.7  \\
7514.1  &  0.45 &  0.07   &   Fe\,{\sc ii} (m73) 7515.9  \\
7712.2  &  0.59 &  0.10   &   Fe\,{\sc ii} [m30F] 7710.8 \\
7751.0  &  14.91 &  2.30   &   Ar\,{\sc iii} [m1F] 7753.2 \\
7775.1  &  0.46 &  0.08   &   O\,{\sc i} (m1) 7775.4 \\
7817.2  &  0.56 &  0.09   &   He\,{\sc i} (m69) 7816.2 \\
7868.9  &  0.42 &  0.05   &   Uid \\
7892.5  &  2.16 &  0.32   &   Uid \\
7917.8  &  0.82 &  0.10   &   Fe\,{\sc ii} [m29F] 7916.9 \\
8131.6  &  0.34 &  0.04  &   Uid \\
8184.4  &  0.38 &  0.05   &   N\,{\sc i} (m2) 8184.8 \\
8309.7  &  0.19 &  0.02   &   Cr\,{\sc ii} [m1F] 8308.7  \\
8317.6  &  0.24 &  0.03   &   Uid \\
8335.0  &  0.33 &  0.04  &   P24 8334.0 \\
8345.7  &  0.50 &  0.05   &   P23 8346.0 \\
8357.5  &  0.67 &  0.08   &   P22 8359.0 \\
8372.9  &  1.22 &  0.14   &   Uid \\
8388.1  &  0.77 &  0.09   &   Uid \\
8406.4  &  0.83 &  0.09  &   Uid \\
8428.5  &  1.24 &  0.12   &   Uid \\
8453.6  &  1.01 &  0.11   &   Uid \\
8462.7  &  10.25 &  1.11   &   Uid \\
8484.2  &  1.67 &  0.15   &   Uid \\
8520.2  &  2.08 &  0.22   &   Uid \\
8566.3  &  1.84 &  0.19   &   Uid \\
8622.1  &  1.71 &  0.17   &   Uid \\
\hline
\end{tabular}
\label{Table_1}
\end{table}

\end{document}